\documentclass[12pt]{elsarticle}
\usepackage{ifluatex}
\usepackage{ifthen}
\newcommand{\grafik}[1]
{
  \ifthenelse{\boolean{true}}{#1}{\includegraphics[width=3cm]{dummy.png}}
}

\usepackage[english]{babel}
\usepackage{shellesc}
\usepackage{graphicx}
\usepackage{xcolor}
\usepackage{soul}
\usepackage{epsfig}
\usepackage{epstopdf}
\usepackage{amssymb}
\usepackage{amsmath}
\usepackage{float}
\usepackage{multirow}
\usepackage{tikz}
\usepackage{pgfplots}
\usepackage{pgfplotstable}
\usepackage{pgffor}
\pgfplotsset{compat=newest}
\usepgfplotslibrary{groupplots}
\usetikzlibrary{pgfplots.groupplots}
\usepackage[T3,T1]{fontenc}
\DeclareSymbolFont{tipa}{T3}{cmr}{m}{n}
\DeclareMathAccent{\invbreve}{\mathalpha}{tipa}{16}
\usepackage{blindtext}
\usepackage[abs]{overpic}
\usepackage{array}
\usepackage{xfrac}
\usepackage{booktabs}
\usepackage{pdflscape}
\usepackage[font=scriptsize,labelfont=bf,singlelinecheck=false,format=plain,,justification=justified,indention=0cm]{caption}
\usepackage[font=scriptsize,labelfont=bf,singlelinecheck=false,format=plain,justification=justified,indention=0cm]{subcaption}
\usepackage[left=1in,right=1in,top=1in,bottom=1in]{geometry}
\usepackage{siunitx}
\renewcommand{\vec}[1]{\mathbf{#1}}
\newcommand{\norm}[1]{\left\lVert#1\right\rVert}

\renewcommand{\b}{\mathrm{b}}

\renewcommand{\c}{\mathrm{c}}
\renewcommand{\d}{\mathrm{d}}

\renewcommand{\i}{\mathrm{i}}

\renewcommand{\l}{\mathrm{l}}

\newcommand{\n}{\mathrm{n}}
\renewcommand{\o}{\mathrm{o}}

\renewcommand{\r}{\mathrm{r}}

\newcommand{\s}{\mathrm{s}}

\renewcommand{\v}{\mathrm{v}}

\newcommand{\w}{\mathrm{w}}

\renewcommand{\sf}{\mathrm{sf}}
\renewcommand{\lg}{\mathrm{lg}}
\newcommand{\slg}{\mathrm{slg}}

\newcommand{\anl}{\mathrm{anl}}

\renewcommand{\min}{\mathrm{min}}
\renewcommand{\max}{\mathrm{max}}

\newcommand{\Bo}{\mathrm{Bo}}

\renewcommand{\Re}{\mathrm{Re}}

\newcommand{\normal}[1]{\hat{\mathbf{#1}}} 
\newcommand{\normalsm}[1]{\hat{\tilde{\mathbf{#1}}}} 

\date{today}

\makeatletter
\def\@author#1{\g@addto@macro\elsauthors{\normalsize%
		\def\baselinestretch{1}%
		\upshape\authorsep#1\unskip\textsuperscript{%
			\ifx\@fnmark\@empty\else\unskip\sep\@fnmark\let\sep=,\fi
			\ifx\@corref\@empty\else\unskip\sep\@corref\let\sep=,\fi
		}%
		\def\authorsep{\unskip,\space}%
		\global\let\@fnmark\@empty
		\global\let\@corref\@empty 
		\global\let\sep\@empty}%
	\@eadauthor={#1}
}
\makeatother

\begin{document}

\begin{frontmatter} 
		\title{Surface tension and wetting at free surfaces\\ in Smoothed Particle Hydrodynamics}
	\author[addr]{Michael Blank\corref{cor}}
	\cortext[cor]{Corresponding author}
	\address[addr]{Institute for Multiscale Simulation, Friedrich-Alexander Universit\"at Erlangen-N\"urnberg, Erlangen, Germany}
	\ead{michael.u.blank@fau.de}
	\author[addr2]{Prapanch Nair}
	\address[addr2]{Department of Chemical Engineering, Indian Institute of Technology Delhi, New Delhi, India}
	\author[addr]{Thorsten P\"oschel}




	\begin{abstract}
Surface tension and wetting are dominating physical effects in micro and nanoscale flows. We present an efficient and reliable model of surface tension and equilibrium contact angles in Smoothed Particle Hydrodynamics for free-surface problems. We demonstrate its robustness and accuracy by simulating several notoriously difficult three-dimensional free surface flow problems driven by interfacial tension.
	\end{abstract}

\end{frontmatter}

\section{Introduction}
\begin{figure}[hb]
  \centering
  \grafik{
\begin{tikzpicture}
	\begin{axis}[
		width = 0.8\linewidth,
		scale only axis,
		xmin=0,
		xmax=857,
		ymin=-100,
		ymax=296,
		axis equal image,
		axis line style={draw=none},
		tick style={draw=none},
		xticklabels={,,},
		yticklabels={,,},
		inner sep=0em,
		]
		\addplot[thick,blue] graphics[xmin=0,ymin=0,xmax=857,ymax=296] {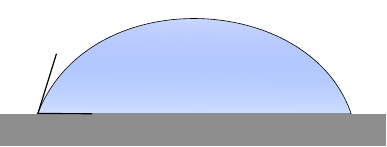};
		
		\node[text width=3cm] at (axis cs:500,35) {solid};
		\node[text width=3cm] at (axis cs:500,190) {liquid};
		\node[text width=3cm] at (axis cs:500,275) {gas};
		
		\node[] at (axis cs:130,90) {$\Theta_{\infty}$};
		
		\draw[-latex] (axis cs:450,-20) to[out=90,in=-90] (axis cs:550,68);
		\node[text width = 4cm] at (axis cs:450,-40) {solid-liquid interface};
		
		\draw[-latex] (axis cs:720,220) to[out=-90,in=33] (axis cs:778,66);
		\node[text width = 3.0cm] at (axis cs:730,260) {three-phase \mbox{contact line}};
		
		\draw[-latex] (axis cs:190,275) to[out=0,in=125] (axis cs:310,245);
		\node[text width = 2.5cm,anchor=west] at (axis cs:0,265) {liquid-gas interface};
		
		\draw[-latex] (axis cs:100,-20) to[out=65,in=225] (axis cs:87,70);
		\node[text width = 3cm] at (axis cs:150,-40) {contact point};
		
		\draw[-latex] (axis cs:720,-20) to[out=90,in=-90] (axis cs:820,66);

		\draw[latex-latex] (axis cs:160,67) to[out=90,in=-33] (axis cs:120,150);
		\node[text width = 3.3cm] at (axis cs:720,-40) {solid-gas interface};
	\end{axis}
      \end{tikzpicture}
      }
	\caption{Sketch of a droplet in contact with a solid substrate in equilibrium}
	\label{fig:Terminology}
\end{figure}
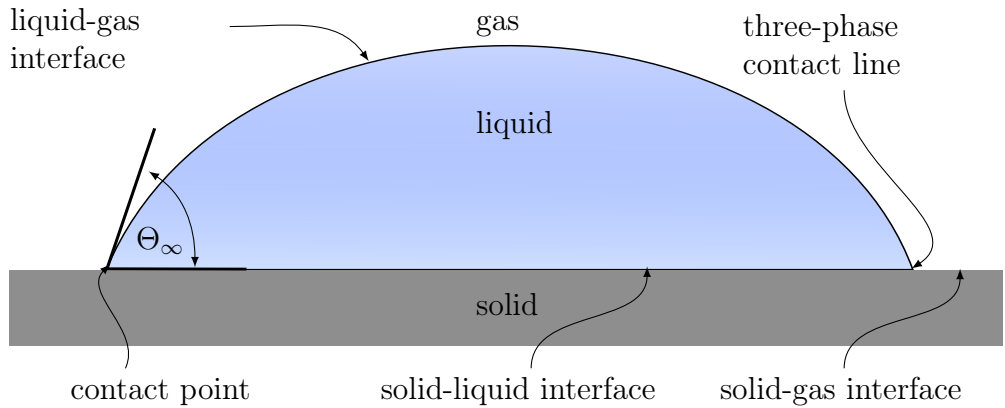

At the micro and nano scale, fluid flows are dominated by interfacial tension that arises at the interface between different phases. When a liquid drop is in contact with a plane solid substrate, the wetting force acting at the triple line leads to an equilibrium contact angle as illustrated in Fig. \ref{fig:Terminology}. 
Surface tension and wetting are essential for many phenomena, including technological processes like oil recovery \cite{Babadagli2002, Babadagli2005}, two-phase heat transfer \cite{Ebadian2011}, and ink-jet printing \cite{Calvert2001}. The wetting properties of a solid surface depend on its chemical composition and morphology, determining its repellent or attractive behavior \cite{Kam_2012, Lai2013}.

In Smoothed Particle Hydrodynamics (SPH) \cite{Lucy1977, GingoldMonaghan1977}, surface tension can be modeled either by pairwise forces between the particles \cite{Nugent2000, Tartakovsky2005, Tartakovsky2016,nair2018} or by a phenomenological Continuum Surface Force (CSF) \cite{Brackbill1992, Morris2000, SIROTKIN2012}.
The concept of pairwise forces between SPH particles is motivated by the molecular origin of surface tension \cite{Rowlinson2013} and readily applies to the simulation of free surfaces. Here, the attractive forces between particles of different phases must be calibrated to obtain a desired equilibrium contact angle. This approach is widely used for solving problems in droplet dynamics, drop interaction with textured surfaces \cite{shigorina2017smoothed}, and contact angle hysteresis \cite{bao2019modified}. When used in simulations, however, such models
 require a large support domain for particles, affecting the computational performance significantly\cite{nair2018}. 


CSF can reliably predict the dynamics due to surface tension gradients at phase boundaries, known as the Marangoni effect. However, its application to free surfaces is problematic since the truncation of the kernel near the free surface leads to unacceptable errors in the local curvature. Incompressible Gas-liquid two phase flow problems can be reliably modelled as free surface flows whenever the shear stress due to the gas phase is negligible. Thus the limitation in modeling surface tension at the free surface has severely restricted the scope of 3D free surface SPH simulations to high Weber number flows. 

This  kernel truncation error at the free surface in CSF has been addressed using several strategies. Mirror particles were employed by Ordoubadi et al. \cite{Ordoubadi2017} to produce robust two-dimensional flow simulations in weakly compressible SPH. Hirschler et al. \cite{Hirschler2017} used CSF to simulate two-dimensional droplet collisions at the free surface using kernel correction parameters. Unfortunately, only two-dimensional systems can be described using any of these approaches. For three dimensional cases, \cite{GEARA2022} introduced an analytical coefficient to account for truncation of kernels near free surface in SPH operators.  The curvature estimate at the free surface was recently improved by F\"urstenau et al. \cite{Fuerstenau2020}, while Blank et al. \cite{Blank2022} developed a method to describe surface tension using a Young-Laplace pressure boundary condition in SPH.

When using CSF, the wetting force in the triple line region can be modeled using the smoothed normal correction scheme introduced by Breinlinger et al. \cite{Breinlinger2013}. Here, the normal vectors (pointing from the liquid into the gas phase), computed at the positions of SPH particles that are located in the vicinity of the three-phase contact line, are modified. Doing so, the curvature is computed at the positions of those SPH particles, and thus, wetting is incorporated into CSF.
This approach can, however, not be applied to free surface flow problems due to the inaccurate force computation caused by insufficient support of particles near the triple contact line region.

The current paper proposes a CSF model that relies on an improved smoothed normal correction scheme. We will show that this model reliably describes three-dimensional free surface problems. We simulate several problems using Incompressible Smoothed Particle Hydrodynamics (ISPH) to demonstrate the model's accuracy and robustness. These problems comprise notoriously difficult systems like droplet oscillation, droplet spreading on a solid substrate, and the pinning of a drop at the contact line of two inclined planes.

\section{Smoothed Particle Hydrodynamics}
Smoothed Particle Hydrodynamics (SPH) is a numerical method for solving continuum problems \cite{Lucy1977, GingoldMonaghan1977, Monaghan2005}. Unlike in mesh-based methods, the domain is discretized by SPH particles with a certain mass at which properties such as velocity, $\vec{u}$,  pressure, $p$, or density $\rho$, are defined. 

The integral interpolant, $\Phi^\mathrm{I}(\vec{r})$, of a physical property, $\Phi(\vec{r})$, at position $\vec{r}$, can be obtained from a spatial convolution with a kernel function, $W(\vec{r}-\vec{r}',h),$ having a compact support:
\begin{equation}
	\label{eq:kernelInterpolation}
	\Phi^\mathrm{I} (\vec{r})=  \int_{\Omega_\c} \Phi(\vec{r}') W(\vec{r}-\vec{r}',h) \d \vec{r}^\prime\,.
\end{equation}
Here, $h$ is the smoothing length, and $\Omega_\c$ denotes the whole continuous domain.   
We approximate the integral in Eq. \eqref{eq:kernelInterpolation} by a sum over
the set $\Omega$ of all SPH particles,
\begin{equation} 
	\Phi^\mathrm{s}\left(\vec{r}\right) = \sum_b \frac{m_b}{\rho_b} \Phi\left(\vec{r}_b\right) W\left(\vec{r}-\vec{r}_b,h \right)\,,
	\label{eq:sumkernel}
\end{equation}
where $\vec{r}_b$, $m_b$, and $\rho_b$ are the position, mass, and density of particle $b$.  In particular, the summation interpolant at the position $\vec{r}_a$ of an SPH particle reads
\begin{equation} 
  \Phi^\mathrm{s}\left(\vec{r}_a\right) = \sum_b \frac{m_b}{\rho_b} \Phi\left(\vec{r}_b\right) W\left(\vec{r}_a-\vec{r}_b,h \right)\,.
  \label{eq:average}
\end{equation}
In the same approximation, the spatial derivative of a physical quantity, $\nabla\Phi\left(\vec{r}\right)$, at position $\vec{r}_a$,  can be expressed in terms of the kernel gradient ${\nabla_{}} W\left(\vec{r}_a-\vec{r}_b,h \right)$ by
\begin{equation}
	\nabla\Phi^\mathrm{s}\left(\vec{r}_a\right) = \sum_b \frac{m_b}{\rho_b} \Phi\left(\vec{r}_b\right) {\nabla_{}} W\left(\vec{r}_a-\vec{r}_b,h \right)\,.
	\label{eq:gradientAverage}
\end{equation}
We introduce shorthand notations and rewrite Eqs. \eqref{eq:average} and \eqref{eq:gradientAverage}: 
\begin{equation}
	\Phi^\mathrm{s}_{a} = \sum_b \frac{m_b}{\rho_b} \Phi_b W_{ab}\,,\qquad \nabla \Phi^{\mathrm{s}}_{a} = \sum_b \frac{m_b}{\rho_b} \Phi_b \nabla W_{ab} \,.
	\label{eq:average_short}
\end{equation}
Here we use the Wendland C$^2$ kernel function normalized for three spatial dimensions \cite{Wendland1995},
\begin{equation}
	\label{WendlandKernel}
	W_{ab}\equiv W\left(\vec{ r}_a - \vec{ r}_b, h\right) = 
	\begin{cases}
		\frac{21}{16\pi} \frac{1}{h^3}\left(1-\frac{r_{ab}}{2h}\right)^4\left(\frac{2r_{ab}}{h}+1\right) & r_{ab} \leq 2h \\
		0 & r_{ab} > 2h
	\end{cases}\,,\qquad r_{ab} \equiv \norm{\vec{ r}_a - \vec{ r}_b}\,.
\end{equation}
We describe the dynamics of a Newtonian incompressible fluid using the Navier-Stokes equation
\begin{equation}
	\label{eq:Momentum}
	\frac{\mathrm{D}\vec{u}}{\mathrm{D} t} = -\frac{1}{\rho}\nabla p + \nu \nabla^{2} \vec{u} + \vec{f}^{\mathrm{s}} + \vec{f}^{\mathrm{b}}\,,
\end{equation}
and the continuity equation
\begin{equation}
  \label{eq:mass}
  \frac{\partial \rho}{ \partial t} = - \nabla\cdot \left(\rho \vec{u}\right)\,,
\end{equation}
where $\vec{u}$, $\rho$, $p$, $\nu$,  are the velocity, density, pressure, and kinematic viscosity, $\vec{f}^{\mathrm{b}}$  is an acceleration due to a body force such as gravity, and $t$ is time. The acceleration caused by surface tension, $\vec{f}^{\mathrm{s}}$, is incorporated into Eq. \eqref{eq:Momentum} as a body force according to the Continuum Surface Force (CSF) model \cite{Brackbill1992}. 

The total acceleration of a particle $a$ given by Eq. \eqref{eq:Momentum} is, thus, a sum of accelerations
\begin{equation}
	\label{eq:NavierStokesIncom}
	\frac{\mathrm{D} \vec{u}_a}{\mathrm{D} t} = \vec{f}^{\mathrm{p}}_a + \vec{f}^{\mathrm{v}}_a + \vec{f}^{\s}_a +  \vec{f}^{\mathrm{b}}_a\,
\end{equation}
due to pressure gradient, $\vec{f}^\mathrm{p}_{a}$, viscosity, $\vec{f}^\mathrm{v}_{a}$, surface tension, $\vec{f}^{\s}_a$, and  body forces  per unit mass, $\vec{f}^{\mathrm{b}}_a$.

The first contribution, the acceleration due to pressure gradient, can be approximated by \cite{Monaghan2005}
\begin{equation}
	\label{eq:PressureGradient}
	\vec{f}^\mathrm{p}_{a} = 
	- \sum_b m_b\left(\frac{p_a }{\rho_a^2} + \frac{p_b }{\rho_b^2}\right) \nabla W_{ab}\,,
\end{equation}
where $p_a \equiv p\left(\vec{ r}_a\right)$, $p_b \equiv p\left(\vec{ r}_b\right)$,  $\rho_a \equiv \rho\left(\vec{ r}_a\right)$, $\rho_b \equiv \rho\left(\vec{ r}_b\right)$. Equation \eqref{eq:PressureGradient} ensures conservation of linear momentum, and
is appropriate for SPH particles, $a$, that are fully embedded by other SPH particles, $b$. This is the case with particles that are in the bulk of the material, far from surfaces. This approximation cannot be applied to particles, $a$, near a free surface. This case will be discussed in Sec. \ref{section:KernelCorrection}.

The second contribution to the acceleration in Eq. \eqref{eq:NavierStokesIncom}, the  viscous acceleration of particle $a$, is given by \cite{Morris1997}
\begin{equation}
	\label{eq:viscousForce}
	\vec{f}^\mathrm{v}_{a} = \sum_b \frac{m_b\left(\eta_a + \eta_b\right) \vec{u}_{ab}}{\rho_a\rho_b} 
	F_{ab}\,,\qquad \vec{u}_{ab} \equiv \vec{u}_a - \vec{u}_b\,,
\end{equation}
where $\vec{u}_a$ and $\vec{u}_b$ are the velocities of particle $a$ and $b$, $\eta_a$ and $\eta_b$ are the dynamic viscosity coefficients assigned to particles $a$ and $b$, and $F_{ab}$ is given by 
\begin{equation}
  F_{ab} \equiv F\left(\vec{r}_{a}-\vec{r}_{b}\right) \textsl{=} \frac{\vec{r}_{ab} \cdot \nabla W_{ab} }{r_{ab}^2 +  \left(0.01h\right)^2}\,, \qquad \vec{r}_{ab}\equiv \vec{r}_a-\vec{r}_b\,.
\end{equation}
Here the term $\left(0.01h\right)^2$ in the denominator avoids divergence of $F_{ab}$ when particles $a$ and $b$ come very close to each other.

The third contribution in Eq. \eqref{eq:NavierStokesIncom}, the acceleration of a particle $a$ due to surface tension, $\vec{f}^{\mathrm{s}}$, will be described in Sec. \ref{section:surfTen}.

\section{Incompressible Smoothed Particle Hydrodynamics (ISPH)}
For incompressible fluids, the velocity field is divergence-free, and the continuity equation, Eq. \eqref{eq:mass}, turns into 
\begin{equation}
  \label{eq:massincompress}
  \nabla \cdot \vec{u} = 0\,.
\end{equation}
The idea of Incompressible Smoothed Particle Hydrodynamics (ISPH) is to enforce Eq. \eqref{eq:massincompress} by imposing a pressure field in such a way that the velocity field becomes divergence-free \cite{Cummins1999}.
In ISPH, we use the predictor-corrector integration scheme by Cummins and Rudman \cite{Cummins1999} to advance particles in space. At time step $n$, the prediction step computes the positions of the SPH particles
from their current positions
and velocities,
\begin{equation}
	\vec{r}_{a}^\ast = \vec{r}_{a}^n + \vec{u}_{a}^n \Delta t\,;\qquad a=1,2,3,\dots
\end{equation}

The corresponding velocities are 
\begin{equation}
  \vec{u}_{a}^\ast = \vec{u}_{a}^n + \left(\vec{f}_{a^\ast}^{\mathrm{v}} + \vec{f}_{a^\ast}^{\mathrm{s}} +\vec{f}_{a^\ast}^{\mathrm{b}}\right)\Delta t
  \quad \text{with}\quad 
  \vec{f}_{a^\ast}^\v \equiv \vec{f}^\v\left(\vec{r}^\ast_a\right)\,,
  \quad \text{etc.}
\end{equation}
The pressure, $p^\ast_a$, corresponding to the predicted velocity can be obtained by solving the pressure Poisson equation (PPE)
\begin{equation}
  \label{PPE0}
  \nabla \cdot \left( \frac{\nabla p_{a}^{\ast}}{\rho_{a}}\right) = \frac{\nabla \cdot \vec{u}_{a}^\ast}{\Delta t}\,.
\end{equation}
From the predicted velocities, $\vec{u}_a^\ast$, and the corresponding pressure, $p^\ast_a$, the correction step yields the divergence-free velocities
at time step $n+1$,
\begin{equation}
	\vec{u}_{a}^{n + 1} = \vec{u}_{a}^\ast - \frac{1}{\rho_{a}}\nabla p_{a}^{\ast}\,\Delta t\,.
\end{equation}
The corresponding particle positions at time step $n+1$ are
\begin{equation}
  \vec{r}_{a}^{n + 1} = \vec{r}_{a}^n + \left(\frac{\vec{u}_{a}^{n} + \vec{u}_{a}^{n + 1}}{2}\right) \Delta t\,.
\end{equation}

The left-hand side of the PPE, Eq. \eqref{PPE0}, can be discretized to obtain \cite{Cummins1999}
\begin{equation}
  \label{eq:PPE_lhs}
  \nabla \cdot \left( \frac{\nabla p_{a}^{\ast}}{\rho_{a}}\right) = 
  \sum_b \frac{m_b}{\rho_b} \frac{4}{\rho_a +
    \rho_b} \left(p_a^{\ast} - p_b^{\ast}\right) F_{ab}\,.
\end{equation}
and its right-hand side results in \cite{Monaghan2005}
\begin{equation}
  \label{eq:PPE_rhs}
  \frac{\nabla \cdot \vec{u}_{a}^{\ast}}{\Delta t} = \sum_b -\frac{m_b}{\rho_b} \frac{\vec{u}^{\ast}_{ab} \cdot
    \nabla W_{ab}}{\Delta t}\,, \qquad \vec{u}^{\ast}_{ab} \equiv \vec{u}^{\ast}_{a} - \vec{u}^{\ast}_{b}\,.
\end{equation}
Substituting Eqs. \eqref{eq:PPE_lhs} and \eqref{eq:PPE_rhs} into Eq. \eqref{PPE0} yields the discretized PPE 
\begin{equation}
  \label{eq:PPE0}
  \sum_b \frac{m_b}{\rho_b} \frac{4}{\rho_a +
    \rho_b} \left(p_a^{\ast} - p_b^{\ast}\right) F_{ab} = \sum_b -\frac{m_b}{\rho_b} \frac{\vec{u}^{\ast}_{ab}\cdot
    \nabla W_{ab}}{\Delta t}\,.
\end{equation}
Equation \eqref{eq:PPE0} can be uniquely solved using any iterative linear solver such as the BiCGSTAB method \cite{Sleijpen1994}, provided the linear system is not singular, e.g., if a Dirichlet boundary condition is applied.
The discretized PPE, Eq. \eqref{eq:PPE0}, is an appropriate approximation of Eq. \eqref{PPE0} for well-embedded SPH particles, that is, for particles in the bulk of the material. The corresponding approximation for SPH particles near a free surface is described in the subsequent section.

\section{Kernel correction for particles near a free surface}
\label{section:KernelCorrection}
In the vicinity of boundaries, the summations in Eqs. \eqref{eq:average} and \eqref{eq:gradientAverage} underestimate $\Phi^\mathrm{I}(\vec{r})$ and $\nabla\Phi^\mathrm{I}(\vec{r})$, due to lacking SPH particles in the neighborhood of $\vec{r}$. We call such SPH particles \textit{insufficiently supported} by neighboring SPH particles. To account for such particles, semi-analytical \cite{Nair2014} or numerical normalization techniques  \cite{Bonet1999, Oger2007}  are used near free boundaries.
In the current paper, akin to \cite{Nair2014} we do not model the gas phase by SPH particles, but we describe the influence of the ambient gas on the liquid through boundary conditions for the liquid phase at the free surface.

To identify SPH particles with insufficient support we use the Shepard filter \cite{Shepard1968},
\begin{equation}
  S_a = \sum_b \frac{m_b}{\rho_b} W_{ab}\,.
\label{eq:shepard}
\end{equation}
However, unlike the approaches described in \cite{Lee2008} where a thin layer of particles comprises the pressure boundary condition (which deteriorates the accuracy and the stability, see \cite{Nair2014, asai2012stabilized}), in the present approach, the pressure at these particles is also solved for.

Particles with insufficient support are close to interfaces, therefore, this method defines the fluid-gas boundary region. Whether a particle $a$ belongs to the subset of particles representing the bulk, $\Omega^{\mathrm{b}}\subset \Omega$, or to the subset representing the boundary, $\Omega^{\mathrm{fs}}\subset \Omega$ is determined by 
\begin{equation}
\label{eq:ShepardThreshold}
\begin{cases}
  a \in \Omega^{\mathrm{fs}} & \text{if}~~~ S_a \leq 0.95 \\
  a \in \Omega^{\mathrm{b}} & \text{else}\,.
\end{cases}
\end{equation}
For well supported particles,  $a\in\Omega^{\mathrm{b}}$, we evaluate Eq. \eqref{eq:PressureGradient} using the pressure gradient approximation \cite{Monaghan2005}; for $a \in \Omega^{\mathrm{fs}}$ we use the approximation given in \cite{Blank2022}:
\begin{equation}
\label{eq:PressureGradientChoice}
\vec{f}^\mathrm{p}_{a} = 
\begin{cases}
- \sum_b m_b\left(\frac{p_a }{\rho_a^2} + \frac{p_b }{\rho_b^2}\right) \nabla W_{ab} & a\in\Omega^{\mathrm{b}}\\[2ex]
-\sum_{b} m_{b} \left( \frac{p_b- p^{\o}_{a} }{\rho_{b}^2} \right) \nabla W_{a{b}}& a\in\Omega^{\mathrm{fs}}\\
\end{cases}\,.
\end{equation}
Here, $p_a^{\mathrm{o}}$ is the external pressure representing the Dirichlet boundary condition of a particle $a$. Since pressure appears in the Navier-Stokes equation only as a gradient, we set $p^{\mathrm{o}}_a = 0$.

In this work, we assume that the motion of the gas phase follows the liquid phase, that is, the relative velocity between the gas and liquid phase ceases. As a consequence, there is no contribution of the gas phase to the viscous acceleration of SPH particles Eq. \eqref{eq:viscousForce}. 
Analogously, using the same assumption the contribution of the gas phase to the  divergence of velocity approximation on the right-hand side of the PPE in Eq. \eqref{eq:PPE0} is zero. 
Therefore, Eqs. \eqref{eq:viscousForce} and \eqref{eq:PPE0} can be used for both, SPH particles located in the vicinity of a free surface, and SPH particles located in the bulk of the simulated material.

To approximate the left-hand side of Eq. \eqref{eq:PPE0} for SPH Particles that are located in the vicinity of a free surface, we use \cite{Nair2014,Blank2022}
\begin{equation}
	\label{PPEDBC}
	\begin{split}
		\left(p_a - p_{a}^{\o}\right) \underbrace{\sum_{b}\frac{m_b}{\rho_b}\frac{4}{\rho_a +
				\rho_b} F_{ab}}_{\beta} & - \sum_{b} \frac{m_{b}}{\rho_{b}}
		\frac{4p_{b}}{\rho_a + \rho_{b}} F_{a{b}} = \\ 
		& \sum_{b} \frac{m_{b}}{\rho_{b}} \left( -\frac{\vec{u}^{\ast}_{a{b}}\cdot \nabla W_{a{b}}}{\Delta t} -
		\frac{4p_{a}^{\o}}{\rho_a + \rho_{b}}F_{a{b}}\right)
	\end{split}\,.
\end{equation}
Note that the term marked by $\beta$ is constant if the mass and volume assigned to the SPH particles is uniform throughout the domain. 

In conclusion, we approximate the PPE by
\begin{equation}
	\begin{cases}
		\text{Eq.} \eqref{eq:PPE0} &\text{if}\quad S_a > 0.95\\
		\text{Eq.} \eqref{PPEDBC}	& \text{if}\quad S_a \leq 0.95\,.
	\end{cases}
\end{equation}
This ambient pressure application can be used to model surface tension by computing the Young-Laplace pressure boundary condition, as shown in \cite{Blank2022}.

\section{Surface tension at free boundaries}
\label{section:surfTen}
The Continuum Surface Force (CSF) model \cite{Brackbill1992} evaluates the acceleration of a particle $a$ due to surface tension by  
\begin{equation}
	\label{eq:surfTenAc_continuous}
	\vec{f}^{\mathrm{s}}_a = \frac{1}{\rho_a}  \vec{F}^{\mathrm{s}}_a\,,
\end{equation}
where $\vec{F}^{\mathrm{s}}_a$ is a force per unit volume  \cite{Morris2000}
\begin{equation}
	\label{eq:CSFGeneral}
 \vec{F}^{\mathrm{s}}_a = \left(2\sigma_a \kappa_a \normal{n}^{\mathrm{}}_a + \nabla^{\mathrm{s}} \sigma_a\right)  \delta^\mathrm{s}_a\,.
\end{equation}
Here, the index $a$ to $\vec{F}^{\mathrm{s}}_a$ and its constituents must be interpreted such that the quantity refers to a local property of the surface, but is assigned to SPH particle $a$. In this sense, $\sigma_a$ is the local coefficient of surface tension of the phase boundary, assigned to particle $a$. In the SPH literature, we speak briefly of ``surface tension of particle $a$''. Similarly, $\kappa_a$, is the local value of the  mean curvature of the surface, assigned to particle $a$. It is termed ``curvature of particle $a$''. Analogously, the local values of the unit normal vector to the interface pointing from phase I to phase II are assigned to the SPH particles. The contribution $\normal{n}^{\mathrm{}}_a$ assigned to particle $a$ is called ``unit normal vector of particle $a$'' and can be computed by 
\begin{equation}
\label{n0}
\vec{n}^{}_a = -\sum_{b}\frac{m_b}{\rho_b} \nabla W_{ab}\,,\qquad    \normal{n}_a = \frac{\vec{n}_a}{\norm{\vec{n}_a}}\,.      
\end{equation}
The same applies to the  surface gradient of the surface tension, $\nabla^{\mathrm{s}} \sigma_a$. For the translation between surface tension into a volumetric force at the position of an SPH particle $a$ we employ the surface delta function, $\delta^{\mathrm{s}}_a$,  that peaks at the interface.

The second term in Eq. \eqref{eq:CSFGeneral} drives fluid flow tangential to the interface due to the surface tension gradient.  In the current paper, we assume constant surface tension. Therefore, the surface tension gradient and, thus, the second term in Eq. \eqref{eq:CSFGeneral} vanish. The first term in Eq. \eqref{eq:CSFGeneral} describes a force (per unit volume) directed perpendicular to the interface. This force counteracts the curvature of the interface and, thus, minimizes the surface area.

We represent the surface delta function of an SPH particle in Eq. \eqref{eq:CSFGeneral} by \cite{Fuerstenau2020}
\begin{equation}
	 \delta^{\s}_a = \norm{\vec{n}_a^{}}\,,\qquad a\in \Omega^\l\,,
\end{equation}
where $\Omega^\l\subset \Omega$ is the subset of SPH particles representing the liquid phase.

  Both the absolute value and the direction of $\vec{n}_a$ obtained from Eq. \eqref{n0} depend sensitively on the positions of neighboring particles $b$, therefore, na\"{\i}ve computation of the normal unit vector vector, $\normal{n}_a \equiv \vec{n}_a/\norm{\vec{n}_a}$, is subject to large fluctuations leading eventually to inaccurate $\vec{F}^{\mathrm{s}}_a$ when used in Eq. \eqref{eq:CSFGeneral}.  Therefore, instead of using the na\"{\i}ve value, we use a smoothed normal vector \cite{Morris1997,Fuerstenau2020}:
\begin{equation}
	\tilde{\vec{n}}_a^{\mathrm{}} = \frac{1}{S_a}\sum_{b} \frac{m_b}{\rho_b} \vec{n}_b^{\mathrm{}} W_{ab}\,.
	\label{n_l_smoothed}
\end{equation}
The smoothing increases the region where particles
contribute to surface tension. Following our previous notation, $\vec{n}_b^{\mathrm{}}$ is a normal vector at the position of particle $b$. 
In the bulk of the fluid, the smoothed normal vectors have small magnitudes and their orientation is of low significance. Therefore, we discard the normal unit vectors of such SPH particles $a\in\Omega^\l$, whose normal vector's magnitude is smaller than a threshold, $\varepsilon^{\mathrm{n}}\in \left[\tfrac{0.1}{h}, \tfrac{0.2}{h}\right]$. In the simulations presented in this paper, we have used $\varepsilon^{\mathrm{n}}=0.01$ throughout.
Subsequently, the smoothed and filtered normal vectors from Eq. \eqref{n_l_smoothed} are normalized by
\begin{equation}
\label{eq:normal_lg}
\normalsm{n}_a = \frac{\tilde{\vec{n}}_a}{\norm{ \tilde{\vec{n}}_a}}\,.
\end{equation}
The set of unit vectors provided by Eq. \eqref{eq:normal_lg}, is used to compute the mean curvature which is needed to evaluate the first term in Eq. \eqref{eq:CSFGeneral} \cite{Monaghan1992, Morris2000}:
\begin{equation}
\label{eq:curvatureSPH}
{\kappa}_a = -\frac{1}{2} \sum_b\frac{m_b}{\rho_b} \left(\normalsm{n}_a^{} - \normalsm{n}_b^{} \right) \cdot \nabla W_{ab}\,. 
\end{equation}
To increase the accuracy of the curvature approximation in Eq. \eqref{eq:curvatureSPH}, we substitute the kernel gradient with $\hat{\nabla}W_{ab} \equiv \vec{C}_a \nabla W_{ab}$ \cite{Bonet1999,Oger2007}, where
\begin{equation}
\vec{C}_a = \left(\sum_{b}\frac{m_b}{\rho_b}\vec{ r}_{ab} \otimes \nabla W_{ab} \right)^{-1}
\end{equation}
is the correction matrix computed from all neighboring SPH particles $b$. This yields the following curvature approximation with $O(h^2)$ convergence \cite{Bonet1999,Oger2007}
\begin{equation}
\label{eq:curvatureSPHCorr}
{\kappa}_a = -\frac{1}{2} \sum_b\frac{m_b}{\rho_b} \left(\normalsm{n}_a^{} - \normalsm{n}_b^{} \right) \cdot  \hat{\nabla} W_{ab}\,.
\end{equation}

Using Eqs. \eqref{eq:CSFGeneral}, \eqref{eq:surfTenAc_continuous},\eqref{eq:normal_lg},  \eqref{eq:curvatureSPHCorr}, and assuming a constant surface tension coefficient in the simulated material, yields the acceleration of a particle $a$ in the normal direction
\begin{equation}
\label{eq:discreteCSF0}
\vec{f}_{a}^{\mathrm{s}} = 2 \frac{\sigma_{a}}{\rho_a} {\kappa}_{a} \normalsm{n}_a \delta_a^\s\,.
\end{equation}
Similar to \cite{Blank2022}, where the Shepard filter is used to increase the robustness and accuracy of the obtained Young-Laplace pressure jump, we modify Eq. \eqref{eq:discreteCSF0} to
\begin{equation}
\label{eq:discreteCSF1}
\vec{f}_{a}^{\mathrm{s}} = \left(1+\frac{1}{S^{\n}_a}\right) \frac{\sigma_{a}}{\rho_a} {\kappa}_{a} \normalsm{n}_a \delta_a^\s\,.
\end{equation}
Here, $S^{\n}_a$ is the Shepard filter computed by
\begin{equation}
\label{eq:ShepardNormals}
S^{\n}_a = \sum_{b\in\Omega^\n} \frac{m_b}{\rho_b} W_{ab}\,,
\end{equation}
where $\Omega^\n \subset \Omega$ is the subset of SPH particles satisfying $\norm{\tilde{\vec{n}}_a} \geq \varepsilon^\n$. The use of the factor $\left( 1+1/S_a^n\right)$ which replaces the theoretical factor of $2$ is empirical and is obtained from  numerical experimentation. 

\section{Wetting forces at free boundaries}
The spreading force per unit length at the three-phase contact line can be described as a function of the contact angle, $\Theta$, and the equilibrium contact angle, $\Theta_{\infty}$, by \cite{Gennes1985}
\begin{equation}
	\label{eq:WettingForce}
  {S}^{\w} =  \sigma \left(\cos \Theta_{\infty} - \cos\Theta \right)\,.
\end{equation}
According to Breinlinger et al. \cite{Breinlinger2013}, Eq. \eqref{eq:WettingForce} can be imposed on SPH particles located in the vicinity of the three-phase contact line by modifying their normal vector orientations. As a result, the curvature is modified, and Eq. \eqref{eq:discreteCSF1} includes the acceleration due to wetting phenomena.
The scheme of \cite{Breinlinger2013} cannot be directly applied to free surface problems since the gas phase is not represented by SPH particles. Therefore, in the following subsections, we describe two alternative approaches for modeling wetting and non-wetting contact for free surface problems.

\subsection{Wetting contact angle, $\Theta_{\infty} \leq 90^\circ $} \label{section:WettingcontactAngles}
To model wetting contact angles (hydrophilic substrates), we compute the normal vectors of those SPH particles located near the three-phase contact line as a function of the required equilibrium contact angle. 
According to \cite{Brackbill1992} the normal vector associated with a desired equilibrium contact angle, $\Theta_{\infty}$, can be computed for a particle $a\in\Omega^\l$ by
\begin{equation}
	\label{eq:BreinlingerCorrectionParticle}
	\normalsm{n}^{\infty}_{a} = \normalsm{t}_a^{\sf} \sin \Theta_{\infty} - \normalsm{n}_a^{\sf} \cos \Theta_{\infty}\,.
\end{equation}
Here, $\normalsm{t}_a^{\sf}$ is the smoothed normalized tangent vector between the solid and fluid phases, and $\normalsm{n}_a^{\sf}$ is the smoothed normalized normal vector pointing from the solid to the fluid phase computed at the position of particle $a$. 
The normal vector, $\normalsm{n}_a^{\sf}$, at the position of particle $a$, is given by
 \begin{equation}
 		\normalsm{n}_a^{\sf} = \frac{\tilde{\vec{n}}_a^{\sf}}{\norm{\tilde{\vec{n}}_a^{\sf}}}\,,
 		\quad
 		\tilde{\vec{n}}_a^{\sf} = \frac{1}{S_a^{\s}} \sum_{b \in \Omega^{\s}} \frac{m_b}{\rho_b} \vec{n}_a^{\sf} W_{ab}\,,
 		\quad 
 		\vec{n}_a^{\sf} = -\sum_{b \in \Omega^{\s}} \frac{m_b}{\rho_b} \nabla W_{ab}\,,
 \end{equation}
where  $\Omega^{\s}\subset\Omega$ is the subset of all SPH particles representing the solid phase, and $S_a^{\s}$ is the Shepard filter computed by
\begin{equation}
S_a^{\s} = \sum_{b \in \Omega^{\s}} \frac{m_b}{\rho_b} W_{ab}\,.
\end{equation}
Analogous to the computation of $\normalsm{n}_a$, we discard the unit normal vectors of SPH particles if $\norm{\tilde{\vec{n}}}_a < \varepsilon^\n$.

The unit tangent vector in Eq. \eqref{eq:BreinlingerCorrectionParticle}, $\normalsm{t}_a^{\sf}$, is computed by 
\begin{equation}
  \tilde{\vec{t}}^{\sf}_{a} = \normalsm{n}^{}_a - \left(\normalsm{n}_a ^{}\cdot \normalsm{n}^{\sf}_a \right)\normalsm{n}^{\sf}_a \,,\qquad \normalsm{t}^{\sf}_{a} = \frac{\tilde{\vec{t}}_{a}^{\sf}}{\norm{\tilde{\vec{t}}_{a}^{\sf}}}\,.
\end{equation}

The normal vectors computed in Eq. \eqref{eq:BreinlingerCorrectionParticle} could be used to replace the normal vectors given by Eq. \eqref{eq:normal_lg} to model wetting.

However, as shown in \cite{Breinlinger2013}, it is preferable to avoid an instantaneous change of the normal vectors of those SPH particles located near the three-phase contact line region. Instead, a  smooth transition from $\normalsm{n}^{}_a$ to $\normalsm{n}^{\infty}_a$ (Eq. \eqref{eq:BreinlingerCorrectionParticle}) depending on a particle's distance to the solid phase (wall) should be applied 
\begin{equation}
	\label{eq:normalCorr1}
	\normalsm{n}_a^{\lg} = \frac{f_a \normalsm{n}_a^{} + \left(1 - f_a\right)\normalsm{n}^{\mathrm{\infty}}_{a}}{\norm{f_a \normalsm{n}_a^{} + \left(1 - f_a\right)\normalsm{n}^{\infty}_{a}}}\,.
\end{equation}
Here, $f_a$ is given by
\begin{equation}
  f_a = 
  \begin{cases}
    0 & \forall \quad d_{a}^{\s} < 0,\\
    d_{a}^{\s}/r_{\max} & \forall \quad 0 \leq d_{a}^{\s} \leq r_\max\\
    1 & \forall \quad d_{a}^{\s} > r_\max
  \end{cases}\,,
  \label{transitionF}
\end{equation}
where  $r_\max$ is the kernel radius ($\r_\max = 2h$ when using the Wendland kernel in Eq. \eqref{WendlandKernel}), and $d_{a}^{\s}$ is the shortest distance of a particle $a\in\Omega^\l$ to the particles $\b \in \Omega^\s$
\begin{equation}
	\label{eq:distancetosolid}
	d_{a}^{\s} = \min \left(\left(\vec{r}_a - \vec{r}_b\right) \cdot \normalsm{n}_{ a}^{\sf}\right) - \Delta x\,.
\end{equation}
Here, $\Delta x$ is the spacing between two SPH particles when arranged on a square lattice, and $\Delta x^3$ is the volume assigned to an SPH particle. The lower limit of $f_a$ ensures that particles that are located closer to the wall than $\Delta x$ are prescribed with $\normalsm{n}_a^\infty$. The upper limit of $f_a$ restricts the normal correction to particles in a range of $r_{\max}$ to the solid phase.

The curvature at the position of a particle $a\in\Omega^\l$ can be computed by 
\begin{equation}
\label{eq:curvatureWettingNaive}
{\kappa}_a = 
-\frac{1}{2}\sum_{b} \frac{m_b}{\rho_b}\left(\normalsm{n}_a^{\lg} - \normalsm{n}_b^{\lg} \right)\hat{\nabla} W_{ab}
\end{equation}

\begin{figure}[htb]
  \centering
  \grafik{
	\begin{tikzpicture}
		\begin{axis}[
			scale only axis,
			xmin=-1,
			xmax=62.745,
			ymin=-2,
			ymax=53.0,
			axis equal image,
			axis line style={draw=none},
			tick style={draw=none},
			xticklabels={,,},
			yticklabels={,,},
			inner sep=0em,
			width= 0.6\linewidth,
			]
			\addplot[thick,blue] graphics[xmin=0,ymin=0,xmax=62.745,ymax=53.0] {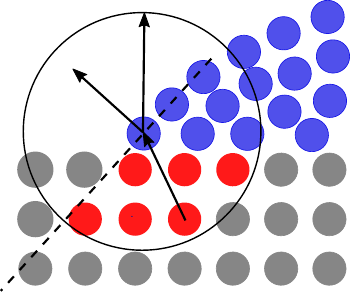};

\node[text width=1.5cm, anchor=west] at (11,43) {$\normalsm{n}_a^{\lg}$};

\node[text width=1.5cm, anchor=west] at (28,48) {$r_{\max}$};
\node[text width=1.5cm, anchor=west] at (21,30) {$a$};
\node[text width=1.5cm, anchor=west] at (28,14) {$b$};
\node[text width=1.5cm, anchor=west] at (28.5,24.5) {$\vec{r}_{ab}$};

\node[text width=3cm, anchor=west] at (-1,11) {\rotatebox{90}{$\vec{r}_{ab}\cdot\normalsm{n}_a^{\lg} < 0$}};
\node[text width=3cm, anchor=west] at (3,0) {$\vec{r}_{ab}\cdot\normalsm{n}_a^{\lg} \geq 0$};

		\end{axis}
              \end{tikzpicture}
              }
	\caption{
		Identification of neighboring particles $b\in\Omega^\s$ which contribute to the curvature computation of a particle $a\in\Omega^\l$. This schematic shows SPH particles near the three-phase contact of a droplet resting on a solid boundary. Gray and blue spheres represent the solid boundary and the liquid phase, respectively. 
		The normal vector $\normalsm{n}_a^{\lg}$ is used to compute the distance of a neighboring particle $b\in\Omega^\s$ to the tangent plane (here shown as a dashed black line) by  $\vec{r}_{ab}\cdot\normalsm{n}_a^{\lg}$. Particles $b\in\Omega^\s$ which satisfy $\vec{r}_{ab}\cdot\normalsm{n}_a^{\lg} \geq 0$ contribute to the curvature computation and are shown by black spheres.
			}
		

\label{fig:identifyParticles}
\end{figure}
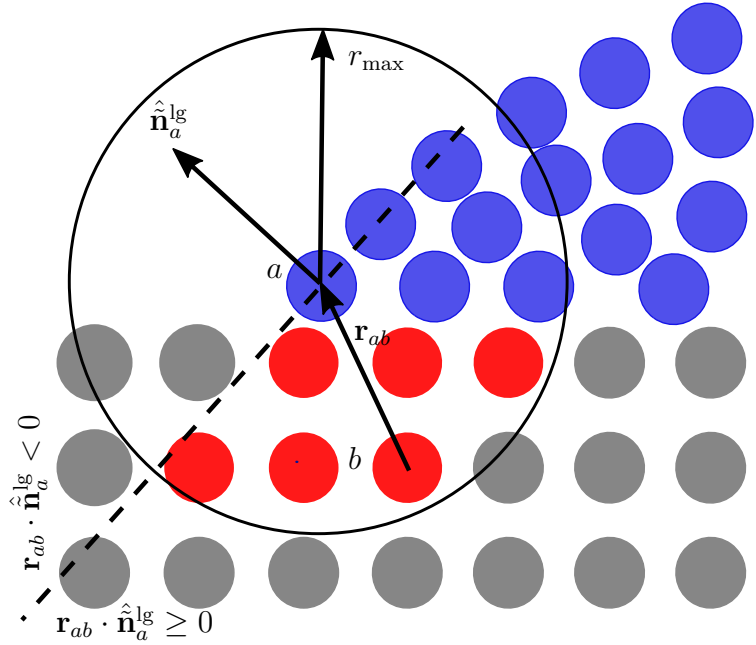

In the course of this work, it was found that computing the curvature in Eq. \eqref{eq:curvatureWettingNaive} is not accurate enough to obtain stable equilibrium droplet shapes for $\Theta_{\infty}\leq 90^{\circ})$. In particular, underestimated curvatures of SPH particles located near the three-phase contact line lead to unlimited spreading of drops on a plane solid substrate.
For this reason, we propose to compute the curvature at the position of an SPH particle $a\in\Omega^\l$ by
\begin{equation}
\label{eq:curvatureWetting}
{\kappa}_a = 
-\frac{1}{2}\sum_{b \in \Omega^{\l}} \frac{m_b}{\rho_b}\left(\normalsm{n}_a^{\lg} - \normalsm{n}_b^{\lg} \right)\hat{\nabla} W_{ab}
-\frac{1}{2}\sum_{b \in \Omega^{\s}} \frac{m_b}{\rho_b}\left(\normalsm{n}_a^{\lg} -	\normalsm{n}^{\s}_{b}\right)\hat{\nabla} W_{ab}\,,
\end{equation}
where $\normal{n}^{\s}_b$ is the normal vector of a neighboring particle $b\in\Omega^\s$ (particles representing the solid phase) given by
\begin{equation}
	\label{eq:modifyselectedSolidNormals}
	\normalsm{n}^{\s}_{b} = 
	\begin{cases}
		\normalsm{t}_a^{\sf} \sin \Theta_{\infty}^{\s} - \normalsm{n}_a^{\sf} \cos\Theta_{\infty}^{\s}\quad & \text{if}~~~\vec{r}_{ab} \cdot \normalsm{n}_a^{\lg} \geq 0\\
		\normalsm{n}_a^{\lg}& \text{else}\,.\\
	\end{cases}
\end{equation}
Here, $\Theta_{\infty}^{\s}$ is a calibration parameter  used to obtain corrected normal vectors of neighboring SPH particles $b\in\Omega^\s$. By setting $\Theta_{\infty}^{\s} > \Theta_{\infty}^{}$, the curvatures at the position of SPH particles $a \in \Omega^\l$ are shifted to larger values, which prevents the continuous spreading of the drop on the solid substrate, which otherwise happens if $\Theta_{\infty}$ is used. This parameter may require calibration for different discretization parameters such as the kernel and the smoothing length. The value of this parameter is listed where relevant, for example, in Table \ref{table:eqCAsSolidLiquid}.
The condition in Eq. \eqref{eq:modifyselectedSolidNormals} measures the distance of a neighboring particle $b\in\Omega^\s$ to particle $a$ in the normal direction. The sign of the distance allows to distinguish between neighboring particles $b\in\Omega^\s$ located on the liquid or on the gas side of particle $a$, as illustrated in Fig. \ref{fig:identifyParticles}.
Neighboring SPH particles that are located on the gas side of particle $a$, $\vec{r}_{ab} \cdot \normalsm{n}_a^{\lg}>0$, do not contribute to the curvature estimate obtained in Eq. \eqref{eq:curvatureWetting}.

\clearpage

\begin{figure}[htb]
	\centering
	\begin{subfigure}[t]{0.45\linewidth}
          \grafik{
          \begin{overpic}[width=\linewidth,trim=100 75 100 125,clip]{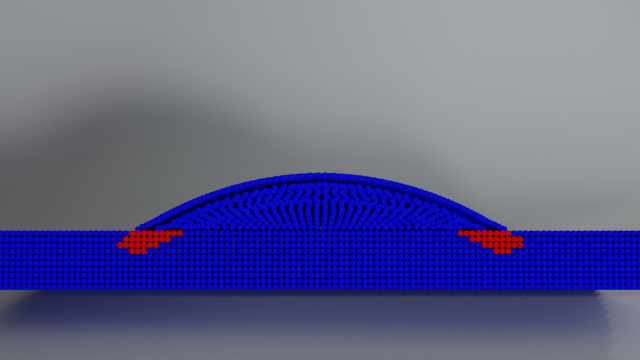}
		\end{overpic}
		\caption{$\Theta_{\infty} = 30^{\circ}$}
	\end{subfigure}
	\quad
	\begin{subfigure}[t]{0.45\linewidth}
		\begin{overpic}[width=\linewidth,trim=100 75 100 125,clip]{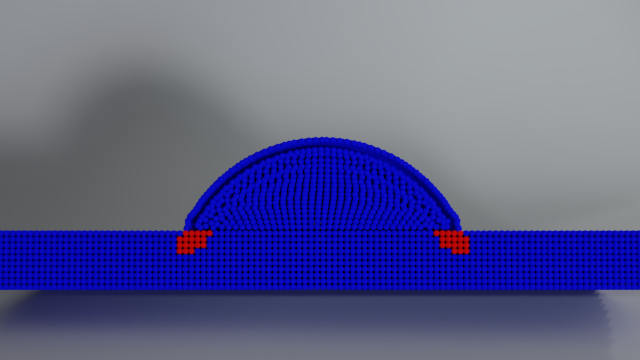}
		\end{overpic}
                }
		\caption{$\Theta_{\infty} = 60^{\circ}$}
	\end{subfigure}
	\caption{Identified solid SPH particles (red) representing an extension of the liquid-gas interface.}
	\label{fig:fixedNormalSolids}
\end{figure}

Figure \ref{fig:fixedNormalSolids} shows cuts through the axis of symmetry of two drops resting on a plane surface. For a desired equilibrium contact angle of $\Theta_{\infty} = 30^{\circ}$ (left) and $\Theta_{\infty} = 60^{\circ}$ (right), the proposed approach modifies the normal vectors of the red-colored SPH particles during the curvature computation. Note that each SPH particle $a \in \Omega^\l$ in the vicinity of the three-phase contact line has its own set of SPH particle neighbors $b \in \Omega^\s$.

The acceleration due to surface tension and wetting experienced by a particle $a\in\Omega^\l$ is now given by 
\begin{equation}
\label{eq:discreteCSF1Wetting}
\vec{f}_{a}^{\mathrm{s}} = \left(1+\frac{1}{S^{\n}_a}\right) \frac{\sigma_{a}}{\rho_a} {\kappa}_{a} \normalsm{n}_a^{\lg} \delta_a^{\s}\,.
\end{equation}
Here, $S^{\n}_a$ is the Shepard filter computed by
\begin{equation}
\label{eq:ShepardNormalsSL}
S^{\n}_a = \sum_{b\in\Omega^{\n}_\l} \frac{m_b}{\rho_b} W_{ab} + \sum_{b\in\Omega^{\n}_\s} \frac{m_b}{\rho_b} W_{ab}\,,
\end{equation}
where $\Omega^{\n}_\l \subset \Omega^\l$ is the subset of SPH particles satisfying 
$\norm{\tilde{\vec{n}}_a} > \varepsilon^\n$ and $\Omega^{\n}_\s \subset \Omega^\s$ is the subset of neighboring SPH particles satisfying 
 $\vec{r}_{ab} \cdot \normalsm{n}_a^{\lg} \geq 0$ in Eq. \eqref{eq:modifyselectedSolidNormals}. 

Using Eqs. \eqref{eq:discreteCSF1Wetting} and \eqref{eq:ShepardNormalsSL}, the acceleration of an SPH particle $a$ due to surface tension and wetting for is now given by  
\begin{equation}
\label{eq:discreteCSFWetting}
\vec{f}_{a}^{\mathrm{s}} = \left(1+\frac{1}{S^{\n}_a}\right) \frac{\sigma_{a}}{\rho_a} {\kappa}_{a} \normalsm{n}_a^{\lg} \delta_a^{\s}\,.
\end{equation}


\subsection{Non-wetting contact angle, $\Theta_{\infty} > 90$}
Instead of computing the curvature using the approach presented in Sec. \ref{section:WettingcontactAngles}, we model non-wetting contact angles by computing the curvature of an SPH particle $a \in \Omega^\l$ using only SPH particle neighbors $b \in \Omega^\l$ representing the liquid phase. In the following, we use the superscript l to denote that a property is computed using only the subset of SPH particles representing the liquid phase.
Analogous to the correction scheme in Sec. \ref{section:WettingcontactAngles}, the  normal vector assigned to an SPH particle $a\in\Omega^\l$ is
\begin{equation}
	\label{eq:BreinlingerCorrectionParticleLiquid}
	\normal{n}^{\l,\infty}_{a} = \normal{t}_a^{\sf,\l} \sin \Theta_{\infty} - \normalsm{n}_a^{\sf} \cos \Theta_{\infty}\,.
\end{equation}
Here, $\vec{t}^{\sf,\l}_{a}$ is the tangent vector at the position of a liquid SPH particle given by
\begin{equation}
	\vec{t}^{\sf,\l}_{a} = \normalsm{n}^{\lg,\l}_a - \left(\normalsm{n}_a ^{\lg,\l}\cdot \normalsm{n}^{\sf}_a \right)\normalsm{n}^{\sf}_a,\qquad \normal{t}^{\sf,\l}_{a} = \frac{\vec{t}_{a}^{\sf,\l}}{\norm{\vec{t}_{a}^{\sf,\l}}}\,, 
\end{equation}
where the normal vector, $\normalsm{n}_a^{\l}$, is computed from liquid SPH particles by
 \begin{equation}
 	\label{eq:normalFromLiquid}
	\normalsm{n}_a^{\l} = 
\begin{cases}
0 & \text{if\ } \norm{\tilde{\vec{n}}_a^{\mathrm{l}}}< \varepsilon^\n\\
\frac{\tilde{\vec{n}}_a^{\mathrm{l}} }{ \norm{ \tilde{\vec{n}}^{\mathrm{l}}_a}} & \text{otherwise\,.}
\end{cases}\,,
	\quad
	\tilde{\vec{n}}_a^{\l} = \frac{1}{S_a^{\l}} \sum_{b \in \Omega^{\l}} \frac{m_b}{\rho_b} \vec{n}_a^{\l} W_{ab}\,,
	\quad
	\vec{n}_a^{\l} = -\sum_{b \in \Omega^{\l}} \frac{m_b}{\rho_b} \nabla W_{ab}\,.
\end{equation}

Here, $S_a^{\l}$ is the Shepard filter applied to liquid SPH particles
\begin{equation}
	\label{eq:ShepardLiquid}
	S_a^{\l} = \sum_{b \in \Omega^{\l}} \frac{m_b}{\rho_b} W_{ab}\,.
\end{equation}
Finally, the normal vector of an SPH particle $a\in\Omega^\l$ located in the vicinity of the three-phase contact line region is given by
\begin{equation}
	\label{eq:normalCorr1l}
	\normalsm{n}_a^{\slg,\l} = 
	\frac{f_a \normalsm{n}_a^{\l} + \left(1 - f_a\right)\normal{n}^{\l,\infty}_{a}}
	{\norm{f_a \normalsm{n}_a^{\l} + \left(1 - f_a\right)\normal{n}^{\l, \infty}_{a}}}\,.
\end{equation}
The modified normal vectors, $\normalsm{n}_a^{\slg,\l}$, replace the normal vectors computed  in Eq. \eqref{eq:normalFromLiquid} if the SPH particle is located in the vicinity of the three-phase contact line 
\begin{equation}
	\label{eq:normalSubstitutionLiquid}
	\normalsm{n}^{\lg}_a = 
	\begin{cases}
		\normalsm{n}^{\slg,\l}_a & 
		\textnormal{if~~} a \in \Omega^{\slg}\\
		\normalsm{n}^{\lg,\l}_a & \textnormal{else\,.}		
	\end{cases}
\end{equation}
Finally, the mean curvature of a liquid SPH particle is given by 
\begin{equation}
	\label{eq:CurvatureLiquidNW}
	{\kappa}_a = - \frac12\sum_{b \in \Omega^{\l}} \frac{m_b}{\rho_b}\left(\normalsm{n}_a^{\lg} - \normalsm{n}_b^{\lg} \right)\hat{\nabla}^{\l} W_{ab}\,,
\end{equation}
where the renormalized kernel gradient, $\hat{\nabla}^{\l}W_{ab}$, is computed from liquid SPH particles $b\in\Omega^\l$ by \cite{Bonet1999,Oger2007}
\begin{equation}
\hat{\nabla}^{\l} W_{ab} \equiv \vec{C}^{\l}_a \nabla W_{ab}\,,\qquad \vec{C}^{\l}_a = \left(\sum_{b \in \Omega^{\l}} \frac{m_b}{\rho_b}\vec{ r}_{ab} \otimes \nabla W_{ab} \right)^{-1}\,.
\end{equation}
Here, $\vec{C}^{\l}_a$ is the correction matrix computed for SPH particles $b \in \Omega^\l$. The acceleration of an SPH particle $a\in\Omega^\l$ due to surface tension and wetting forces is then given by
\begin{equation}
\label{eq:discreteCSF2}
\vec{f}_{a}^{\mathrm{s}} = \left(1+\frac{1}{S^{\n}_a}\right) \frac{\sigma_{a}}{\rho_a} {\kappa}_{a} \normalsm{n}^{\lg}_a \delta^{\s,\l}_a\,,
\end{equation}
with
\begin{equation}
\label{eq:ShepardNormals2}
S^{\n}_a = \sum_{b \in \Omega^\n_\l} \frac{m_b}{\rho_b} W_{ab}\,,\qquad \delta^{\s,\l}_a= \norm{\vec{n}_a^\l}\,.
\end{equation}
For desired equilibrium contact angles approaching $90^\circ$, a slight discrepancy between the wetting and the non-wetting approach is observed, with the wetting approach yielding greater accuracy. 
\subsection{Extreme wetting and thin films}
Obtaining stable drops on a plane surface with $\Theta_{\infty} \leq 30^{\circ}$ is challenging because the particles near the three-phase contact line have much fewer neighbors within their support radius ($b\in\Omega^\l$) than for larger contact angles.  Further, for low contact angles, the resulting sharp curvatures cannot be estimated accurately due to the smoothing of the normals (see Eqs. \eqref{eq:normalSubstitutionLiquid} and \eqref{eq:modifyselectedSolidNormals}). When the liquid phase spreads thinner than the SPH kernel radius
on the substrate, the aforementioned approach will result in all particles being identified as near the contact line because all particles will be near the free surface and the substrate. Consequently, the normal correction in Eqs. \eqref{eq:normal_lg} and \eqref{eq:normalFromLiquid}  will be applied to all particles $a \in \Omega^\l$ in the liquid layer, resulting in unphysical surface forces.

For the case of small contact angle, as simulated in Sec. \ref{sec:sessile}, we substitute the curvature computation in Eq. \eqref{eq:curvatureWetting} by Eq. \eqref{eq:CurvatureLiquidNW} (where a renormalized kernel gradient is used) for an SPH particle $a\in\Omega^\slg$  if there is an insufficient neighbor particle support which we identify by $S_a < 0.6$.

For a thin film of liquid, the thickness of the film may be less than the kernel support radius. This leads to incorrect identification of particles. However, these particles are not necessarily near the triple contact line, but their surface normals can be parallel to the substrate normals.  Therefore, we use the threshold   $\normalsm{n}_a^{\lg} \cdot\normalsm{n}_a^{\sf} \leq 0.995$, for the normal correction (Eq. \ref{eq:normalCorr1}) of particles near the triple contact line to avoid the particles in a thin film being classified as located  near the triple line.

\section{Wall boundary}
To model wetting phenomena on an impermeable substrate using SPH, the wall must satisfy the no penetration and the no-slip boundary conditions \cite{Violeau2015}. 
Several techniques were proposed to ensure no penetration. For instance, wall boundaries can be modeled using repulsive forces \cite{monaghan1994simulating}, dummy particles \cite{crespo2007},  mirror particles \cite{Morris1997}, immersed boundary methods \cite{nasar2019}, or semi-analytical approaches \cite{leroy2014unified}.
To simultaneously satisfy no-penetration and no-slip boundary conditions at walls, we modify the PPE in Eq. \eqref{eq:PPE_lhs} following \cite{adami2012}. We define the set of wall particles  $\Omega^\w$ as a subset of solid particles, $\Omega^\w\subset\Omega^\s$. Wall particles, $a\in\Omega^\w$, are thus near fluid particles.  In mathematical terms,  an SPH particle represents the solid wall boundary, $a\in\Omega^\w$, if 
\begin{equation}
 a\in\Omega^\s\ \text{~~and~~ }S^{\mathrm{l}}_a > 10^{-3}\,.
\end{equation}
The threshold $S^{\mathrm{l}}_a > 10^{-3}$ must be chosen large enough to allow the convergence of the iterative solver solving the PPE. The PPE for particles  representing the wall ($a\in\Omega^\w$) is 
\begin{equation}
	\label{eq:PPEWall}
	\quad \sum_{b \in \Omega^{\l}} \frac{m_{b}}{\rho_{b}} \frac{4}{\rho_{a} +
		\rho_{b}} \left(p_{a}^{\ast} - p_{b}^{\ast}\right) F_{ab} = \sum_{b\in \Omega^{\l}}-\frac{m_{b}}{\rho_{b}} \frac{\vec{u}_{ab}^\ast \cdot
		\nabla W_{ab}}{\Delta t} 
\end{equation}
Thus, the neighborhood of wall particles includes exclusively fluid particles. 
In summary, the pressure of a particle $a$ follows from different PPEs depending on whether it is a liquid particle in the bulk of the fluid, a liquid particle near the free surface, or a wall particle:
\begin{equation}
\begin{cases}
\text{Eq.} \eqref{eq:PPE0} &\text{if}\quad a \notin \Omega^\mathrm{w} \text{ and } S_a > 0.95\\
\text{Eq.} \eqref{PPEDBC}	& \text{if}\quad a \notin \Omega^\mathrm{w} \text{ and } S_a \leq 0.95\\
\text{Eq.} \eqref{eq:PPEWall} &\text{if}\quad a \in \Omega^\mathrm{w}\,.
\end{cases}
\end{equation}
To enforce the no-slip boundary condition at the wall, the wall's velocity is calculated via an SPH interpolation of the liquid neighborhood of $a\in\omega^\mathrm{w}$ as
\begin{equation}
	\tilde{\vec{u}}_{a}^{} = \frac{1}{S_a^{\l}}\sum_{b \in\Omega^{\l}} \frac{m_b}{\rho_b}\vec{u}_{b} W_{ab}\,.
\end{equation}
Subsequently, the wall particle velocity, ${\vec{u}}^{\w}_a$, that replaces the velocity of neighboring particles $b\in\Omega^\mathrm{w}$ in Eq. \eqref{eq:viscousForce}, is given by
\begin{equation}
  \label{eq:WallVelocity}
  {\vec{u}}^{\w}_a = 2 \vec{u}_{a} - \tilde{\vec{u}}_{a}^{}\,.
\end{equation}

\section{Validation of the proposed CSF and wetting model}

\subsection{Numerical parameters}

Table \ref{tab:SimParamValid} summarizes the numerical parameters used in all tests unless otherwise stated. The parameter chosen, albeit arbitrary, belong to the realm of realistic fluids. Stable simulations with properties of water may also be achieved with a higher spatial resolution than provided.   
\begin{table}[htb]
  \centering
  \caption{Numerical parameters used in the test simulations}
  \label{tab:SimParamValid}
  \begin{tabular}{ lll}
    \toprule
    parameter & symbol & value  \\
    \midrule
    surface tension & $\sigma$	& $0.01\,\si{\newton\per\metre}$\\
    dynamic viscosity & $\eta$	& $0.05\,\si{\newton\per\metre}$\\
    mass density & $\rho$ & $1000\,\si{\kg\per\cubic\metre}$ \\
    ambient pressure & $p^\mathrm{amb}$ & 0 \\
    smoothing length & $h$ & $2\Delta x$ \\
    threshold for significance of vectors, see Eq.
              \eqref{eq:normalFromLiquid} & $\varepsilon^{\n}$ & $0.1/h$ \\
    spacing of SPH particles when arranged on a square lattice & $\Delta x$ &  $5\times 10^{-5}\,\si{\metre}$\\
    \bottomrule
  \end{tabular}
\end{table}
The integration time step was chosen due to the Courant-Friedrich-Lewy condition \cite{Morris2000} from the maximum acceleration $\vec{a}_\mathrm{max}$, the maximum velocity $\vec{u}_\mathrm{max}$, the viscous diffusion, and the surface tension:
\begin{equation}
  \Delta t = \frac14\mathrm{min}\left(\sqrt{\frac{h}{\norm{\vec{a}_\mathrm{max}}}},\ \frac{h}{\norm{\vec{u}_\mathrm{max}}},\ 
    \frac{h^2 \rho}{2\eta},\ 
    \sqrt{\frac{h^3 \rho}{2\pi \sigma}}
  \right)\,.
\end{equation}
The solid SPH particles (including the wall particles) are kept at zero velocity throughout the simulations.

\subsection{Test case: Plane Poiseuille flow}

We simulate plane Poiseuille flow between two stationary, infinite planes, located at $x = \pm L$ with  $L=0.5\,\si{\metre}$. The fluid is accelerated in the $y$-direction by the body force $g_y$. The analytical solution for the $y$-component  of  the time-dependent velocity, $u_y$, as a function of the $x$-position reads \cite{Morris1997}
\begin{equation}
u_y(x,t)  = \frac{g_y}{2\nu} x\left(x-L\right) + \sum_{n=0}^{\infty} \frac{4g_y L^2}{\nu \pi^3\left(2n+1\right)^3}
 \sin\left(\frac{\pi x}{L}\left(2n +1\right)\right)\exp\left(-\frac{\left(2n+1\right)^2 \pi^2 \nu}{L^2} t\right)
\label{eq:PoiseuilleFlowAnl}
\end{equation}
In the simulation, we apply periodic boundary conditions in $y$- and $z$-directions, thus, in this example, there is no free boundary. Consequently, to solve the PPE, we apply a Dirichlet boundary condition for the pressure to the solid SPH particles, which are not identified as wall SPH particles. The pressure of the solid phase is computed implicitly by solving
\begin{equation}
	\begin{cases}
		\text{Eq.} \eqref{eq:PPEWall} &\text{if}\quad a \in \Omega^{\w} \\
		\text{Eq.} \eqref{PPEDBC}	& \text{if}\quad   a \in \Omega^{\mathrm{fs}} \land a \notin \Omega^{\w}\\
		\text{Eq.} \eqref{eq:PPE0} & \text{else\,.}
	\end{cases}
\end{equation}
In this validation example, we disregard surface tension effects, hence the momentum balance of the fluid reads
\begin{equation}
	\label{eq:NavierStokesIncomValidWall}
	\frac{\mathrm{D} \vec{u}_a}{\mathrm{D} t} = \vec{f}^{\mathrm{p}}_a + \vec{f}^{\mathrm{v}}_a +  \vec{f}^{\mathrm{b}}_a\,.
\end{equation}
The flow is characterized by the Reynolds number, ${\Re = 2 L u_{y}^{\max}/\nu = 125}$. 

The SPH particles representing the liquid phase are initially placed on a rectangular lattice in the three-dimensional interval $\vec{x} \in \left(-L,\,L\right)^3$.  At ${x = \pm L}$, the boundary is modeled by SPH wall particles, placed at ${L\leq x \leq L + 2 r_{\max}}$  and  ${L - 2r_{\max} \leq x\leq L}$. The thickness of the walls is twice the Wendland kernel radius,  $r_{\max} = 2h$. This is required to apply the zero-pressure Dirichlet boundary condition in Eq. \eqref{PPEDBC}  to solid SPH particles, in order to solve the PPE. The liquid phase is represented by ${L/\Delta x\in \{10,20,40\}}$ SPH particles in each spatial dimension leading to a total of \{8\,000,\ 64\,000,\ 512\,000\} liquid SPH particles in the simulation.

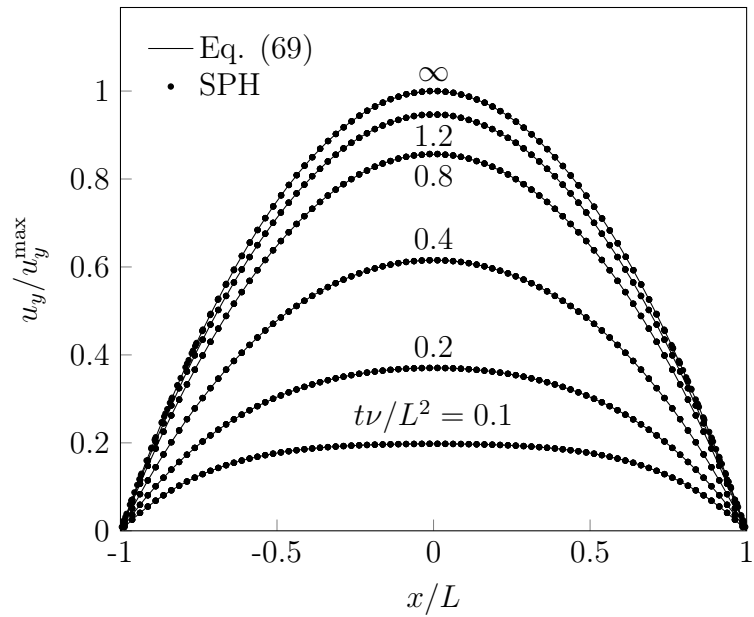
\begin{figure}[htb]
	\centering
        \grafik{
	\begin{tikzpicture}
		\begin{axis}[
			xlabel = $x/L$,
			ylabel =$u_y/u_{y}^{\mathrm{max}}$,
			width = 0.6\linewidth,
			xtick pos=left,
			ytick	pos=left,
			ymax=1.19,
			ymin = 0,
			xmin = -1,
			xmax = 1,
			xtick={-1,-0.5,0,0.5,1},
			xticklabels={-1,-0.5,0,0.5,1},
			legend style={draw=none, fill=none},
			legend pos = north west,
			legend cell align={left},
			]
			\addplot[ draw=black] table [x expr=\thisrowno{0}/0.5,y expr=\thisrowno{1}/1.25 ,col sep=comma] {data/poiseuille/anl/poiseuille_anl_time_2.5s.txt} node[above, pos=0.5]{$t\nu/L^2 = 0.1$};
			
			\addplot[draw=black,forget plot] table [x expr=\thisrowno{0}/0.5,y expr=\thisrowno{1}/1.25 ,col sep=comma] {data/poiseuille/anl/poiseuille_anl_time_5.0s.txt} node[above, pos=0.5]{$0.2$};
			
			\addplot[ draw=black,forget plot] table [x expr=\thisrowno{0}/0.5,y expr=\thisrowno{1}/1.25 ,col sep=comma] {data/poiseuille/anl/poiseuille_anl_time_10.0s.txt} node[above, pos=0.5]{$0.4$};
			
			\addplot[ draw=black,forget plot] table [x expr=\thisrowno{0}/0.5,y expr=\thisrowno{1}/1.25 ,col sep=comma] {data/poiseuille/anl/poiseuille_anl_time_20.0s.txt} node[below, pos=0.5]{$0.8$};
			
			\addplot[ draw=black,forget plot] table [x expr=\thisrowno{0}/0.5,y expr=\thisrowno{1}/1.25 ,col sep=comma] {data/poiseuille/anl/poiseuille_anl_time_30.0s.txt} node[below, pos=0.5]{$1.2$};
			
			\addplot[ draw=black,forget plot] table [x expr=\thisrowno{0}/0.5,y expr=\thisrowno{1}/1.25 ,col sep=comma] {data/poiseuille/anl/poiseuille_anl_time_100.0s.txt} node[above, pos=0.5]{$\infty$};
			
			\addplot[ mark size = 1pt, draw=black, only marks] table [x expr=\thisrowno{0}/0.5,y expr=\thisrowno{1}/1.25 ,col sep=comma] {data/poiseuille/dx0.0125/poiseuille_dx_0.0125_time_2.5s.txt};
			\addplot[mark size = 1pt, draw=black, only marks,forget plot] table [x expr=\thisrowno{0}/0.5,y expr=\thisrowno{1}/1.25 ,col sep=comma] {data/poiseuille/dx0.0125/poiseuille_dx_0.0125_time_5.0s.txt};
			\addplot[ mark size = 1pt, draw=black, only marks,forget plot] table [x expr=\thisrowno{0}/0.5,y expr=\thisrowno{1}/1.25 ,col sep=comma] {data/poiseuille/dx0.0125/poiseuille_dx_0.0125_time_10.0s.txt};
			\addplot[mark size = 1pt, draw=black, only marks,forget plot] table [x expr=\thisrowno{0}/0.5,y expr=\thisrowno{1}/1.25 ,col sep=comma] {data/poiseuille/dx0.0125/poiseuille_dx_0.0125_time_20.0s.txt};
			\addplot[ mark size = 1pt, draw=black, only marks,forget plot] table [x expr=\thisrowno{0}/0.5,y expr=\thisrowno{1}/1.25 ,col sep=comma] {data/poiseuille/dx0.0125/poiseuille_dx_0.0125_time_30.0s.txt};
			\addplot[mark size = 1pt, draw=black, only marks,forget plot,each nth point={1}] table [x expr=\thisrowno{0}/0.5,y expr=\thisrowno{1}/1.25 ,col sep=comma] {data/poiseuille/dx0.0125/poiseuille_dx_0.0125_time_100.0s.txt};
			
			\addlegendentry{Eq. \eqref{eq:PoiseuilleFlowAnl} };			
			\addlegendentry{SPH};
		\end{axis}
              \end{tikzpicture}
              }
	      \caption{Instantaneous velocity profiles for viscous fluid flow between two parallel plates driven by the pressure gradient: the fluid flows in the $y$-direction and its non-dimensional velocity is plotted along the normal direction to the plates.  }
	\label{fig:ValidViscoPoiseuille}
\end{figure}

Figure \ref{fig:ValidViscoPoiseuille} compares the analytical solution, Eq. \eqref{eq:PoiseuilleFlowAnl},  of the transient velocity field with the numerical result for ${L/\Delta x = 40}$. 
To evaluate the precision of the numerical model, we compute the relative deviation of the numerical SPH particles' velocities from the corresponding analytical values for SPH particles located in the interval  $-\Delta x \leq y \leq \Delta x$ and $-\Delta x\leq z \leq \Delta x$, as a function of their $x$-position. For $t\nu/L^2 = 4\,$ for $L/\Delta x = 40$ we obtain the mean relative error $1.6\,\%$. Reducing the spatial discretization to $L/\Delta x = 10$ we obtain the mean relative error $3.8\,\%$.

\subsection{Test case: Laplace pressure jump}
At equilibrium, the pressure inside a drop exceeds the ambient pressure due to the surface tension. This effect is termed the Laplace pressure jump.

In this simulation, we arrange $N$ SPH particles on a rectangular lattice inside a sphere of radius $r_0 = 10^{-3}\,\si{\metre}$, using the discretization given in Tab. \ref{tab:ParamValid1}. All smoothed normal vectors of magnitude smaller than ${\varepsilon^{\n} = \tfrac{0.3}{h}}$ are discarded from the computation of the curvature.
\begin{table}[htb]
	\centering
	\caption{Number of SPH particles and discretization spacing used to validate the Young-Laplace pressure boundary condition.}
	\label{tab:ParamValid1}
	\begin{tabular}{ r l}
		\toprule
		$N$ & $\Delta x$ \\
		\midrule
		$4\,224$	 &${r_0}/{10}$ \\
		$9\,915$	 & ${r_0}/{13.\Bar{3}}$\\
		$14\,321$	 & ${r_0}/{15}$\\
		$113\,104$	 & ${r_0}/{30}$\\
		$268\,096$ 	 &${r_0}/{40}$\\
		\bottomrule
	\end{tabular}
\end{table}

According to the Young-Laplace equation, the pressure drop across a liquid-gas interface at equilibrium is given by \cite{Mugele2019}
\begin{equation}
	\label{eq:YLSteadyState}
	\Delta p^{\mathrm{YL}} = 2\sigma\kappa\,.
\end{equation}
For the parameters given above, we find that the pressure inside the drop, $p_\i$, exceeds the ambient pressure by $20\,\si{\pascal}$ according to 
\begin{equation}
	\label{eq:anlLPJ}
	p^\i = p^\mathrm{amb} + \frac{2\sigma }{r_0}\,.
\end{equation}

\begin{figure}[htb]
	\centering 
	\begin{subfigure}[t]{0.45\linewidth}
		\centering
                \grafik{
		\begin{tikzpicture}
			\begin{axis}[
				width = \linewidth,
				xlabel = $x/r_0$,
				ylabel = $p^{\i}/\Delta p^{\mathrm{YL}}$,
				y label style={yshift=-.3em},
				xtick pos=left, 
				ytick pos=left, 
				ymin = 0.0,
				legend cell align={left}, 
				legend style={at={(axis cs:-0.9,0.0)},anchor=south west, draw=none},			
				cycle list name=color list,
				reverse legend,
				clip mode=individual
				]		
				\addplot[ color=red, mark repeat=1, mark=*, only marks,mark options={scale=0.75} ] table [x expr=\thisrowno{0}/0.001, y expr=\thisrowno{3}/20 ,col sep=comma]{data/pressureDrop_CSF/validSTPress_dx2.5e-5.txt};	
				\addplot[ color=blue, mark repeat=1, mark=*, only marks,mark options={scale=0.75} ] table [x expr=\thisrowno{0}/0.001, y expr=\thisrowno{3}/20 ,col sep=comma]{data/pressureDrop_CSF/validSTPress_dx6.6e-5.txt};	
				\addplot+[draw=black, mark repeat=1, mark=*, only marks,mark options={fill=black,scale=0.75} ] table [x expr=\thisrowno{0}/0.001, y expr=\thisrowno{3}/20,col sep=comma]{data/pressureDrop_CSF/validSTPress_dx1e-4.txt};
				
				\draw [dashed,thick] (axis cs:-1,1.0) -- (axis cs:1,1.0);
				\draw [dashed, thick] (axis cs:-1,1.0) --(axis cs:-1,0.0);
				\draw [dashed, thick] (axis cs:1,1.0) --(axis cs:1,0.0);
				\addlegendentry{$r_0/\Delta x = 40$ };	
				\addlegendentry{$r_0/\Delta x = 15$ };	
				\addlegendentry{$r_0/\Delta x = 10$ };
			\end{axis}
                      \end{tikzpicture}
                      }
		\caption{Pressure profile along the $x$-axis ($y/r_0 = z/r_0 =0$) for a drop centered at $(x,y,z)=(0,0,0)$ for different spatial resolutions. The dashed line shows the analytical solution, Eq. \eqref{eq:YLSteadyState}}
		\label{fig:pressureDropCSFa}
	\end{subfigure}\quad
	\begin{subfigure}[t]{0.45\linewidth}
		\centering
                \grafik{
		\begin{tikzpicture}
			\begin{axis}[
				width = \linewidth,
				xlabel = $r_0/\Delta x$,
				ylabel =$\text{relative error\,/\,}\%$,
				xtick pos=left, 
				ytick pos=left, 
				legend cell align={left}, 
				legend style={draw=none}, 
				]
				\addplot+[mark size = 2.0pt, color=black, mark repeat=1, mark=*, only marks ,mark options={fill=black} ] table [x expr=0.001/\thisrowno{0}, y index=4 ,col sep=comma]{data/pressureDrop_CSF/relErr_pressure2.txt};
			\end{axis}
                      \end{tikzpicture}
                      }
		\caption{Relative deviation of the simulated equilibrium pressure inside the droplet from the analytical value, Eq. \eqref{eq:YLSteadyState}.}
		\label{fig:pressureDropCSFb}
	\end{subfigure}
	\caption{Comparison of the computed equilibrium pressure inside a droplet with the analytical solution of the Young-Laplace equation, Eq. \eqref{eq:YLSteadyState}. The simulation result is averaged over the time interval $0.8 \leq t \leq 1 \,\si{\second}$.}
	\label{fig:pressureDropCSF}
\end{figure}
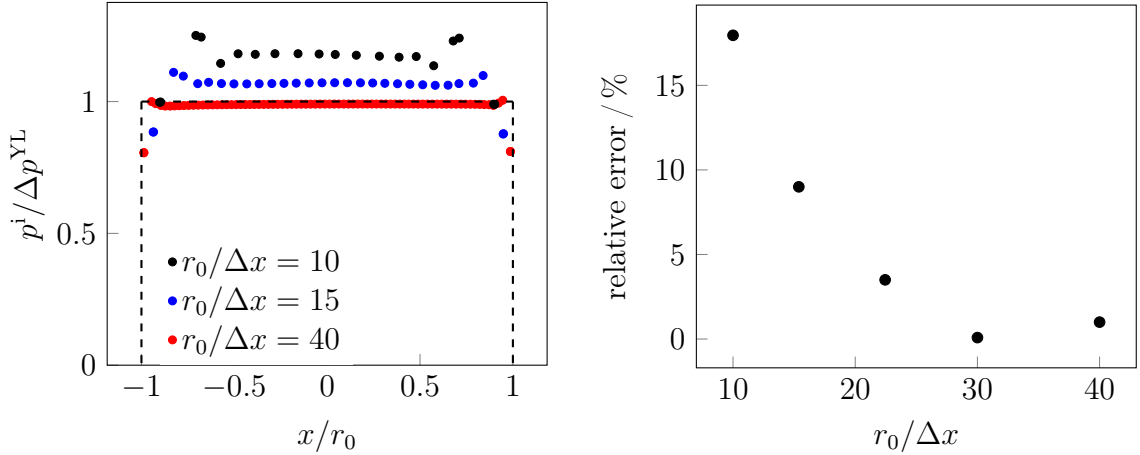

Figure \ref{fig:pressureDropCSF} shows the equilibrium pressure along a cut through the symmetry axis of the drop after a sufficient relaxation time $(t = 1\,\mathrm{s})$ when equilibrium was achieved. For increasing number of SPH particles (increasing resolution), the equilibrium pressure inside the droplet converges to the analytical solution, given by Eq. \eqref{eq:anlLPJ}.
The relative error is $18.0\,\%$ when using 4\,224 SPH particles,  $0.8\,\%$ using 113\,104 SPH particles, and $1.0\,\%$ when 268\,096 SPH particles are used to discretize the droplet.

\subsection{Test case: Droplet oscillations}
\begin{figure}[htb]
	\centering
        \grafik{
          \includegraphics[width=3in,trim=200 175 200 175,clip]{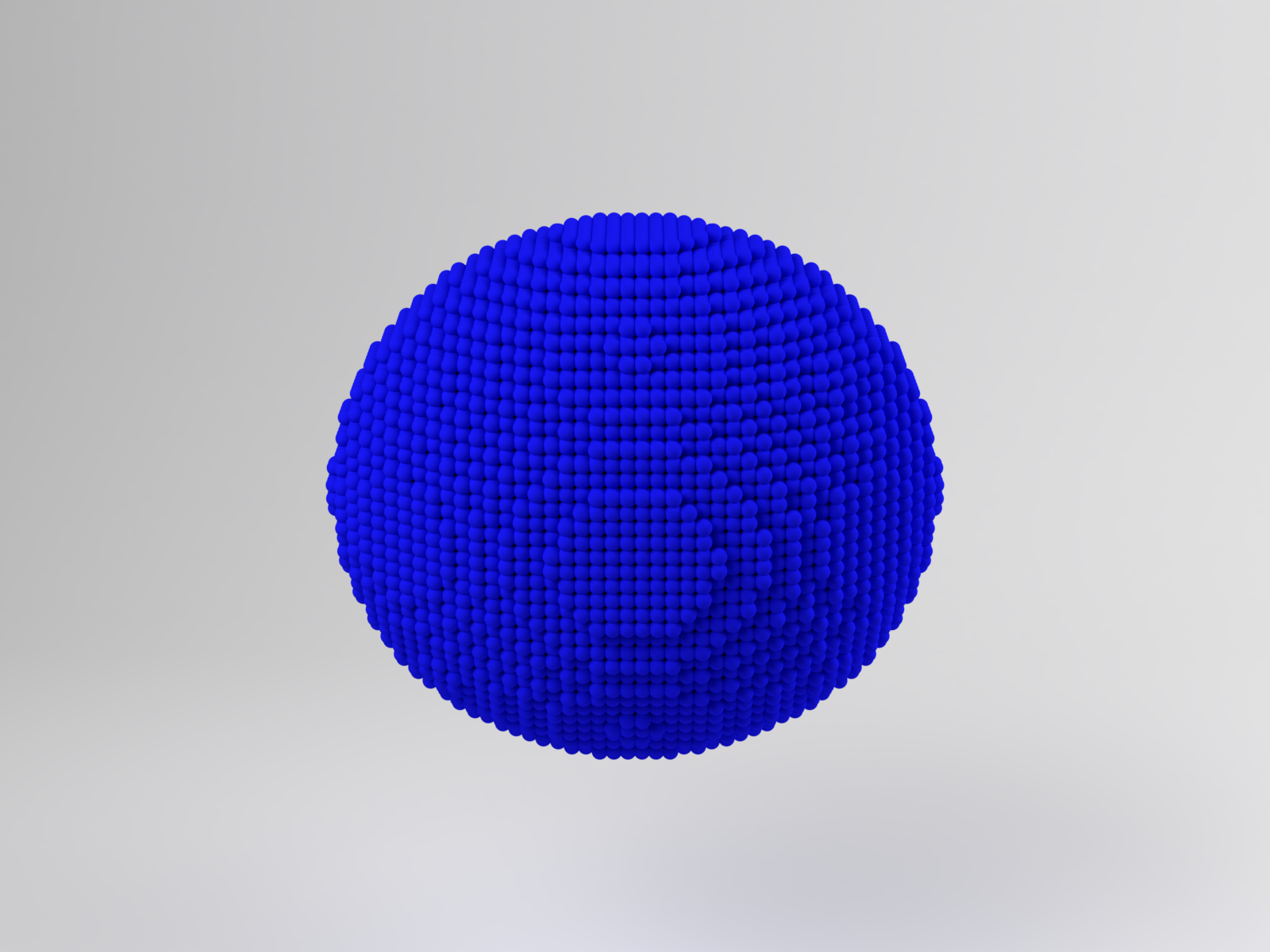}
          }
	\caption{Ellipsoid according to Eq.  \eqref{eq:initshape},  represented by 33\,240 SPH particles.} 
	\label{fig:initialShapeOscillationCSF} 
\end{figure}

In this validation case, we study the damped oscillation of a viscous drop. The analytical solution for the shape of a drop as a function of time reads \cite{Aalilija2020}
\begin{equation}
	r\left(\theta, t\right) = r_0 \left(1+\varepsilon_2(t) P_2\left( \cos\left(\theta\right)\right)-\frac{1}{5}\varepsilon^{2}_2(t) \right)\,.
\end{equation}
Here, $\theta$ is the polar angle,  $r_0$ is the radius of the (spherical) droplet in equilibrium, and $P_2(x)\equiv \frac12\left(3x^2-1\right)$ is the second-order Legendre polynomial. The governing parameter
\begin{equation}
	\varepsilon_2(t) \approx 0.08 \mathrm{e}^{-\lambda_2 t} \cos\left(\omega_{2,0} t\right)\,
\end{equation}
contains the oscillation frequency and the damping constant,
\begin{equation}
	\label{eq:omegaLambda}
	\omega_{2,0} \equiv \sqrt{\frac{8\sigma}{\rho r_0^{3}}}\,,\qquad \lambda_2 \equiv \frac{5\eta}{\rho r_0^2}\,.
\end{equation}
For the initial condition, we assume \cite{Aalilija2020}
\begin{equation}
	r\left(\theta, t = 0\right) = r_0 \left(1+0.08 P_2\left(\cos\left(\theta\right)\right)-\frac{1}{5}0.08^2\right)\,,
 \label{eq:initshape}
\end{equation}
which describes a weakly deformed sphere. For the simulation, we assume $\eta = 5\times 10^{-3}\,\si{\pascal \second}$ and the values given in Tab. \ref{tab:SimParamValid}.
For initialization, the shape described in Eq. \eqref{eq:initshape} is represented by 33\,240 SPH particles placed on a square lattice, see Fig. \ref{fig:initialShapeOscillationCSF}.

\begin{figure}[htb]
	\centering 
  \grafik{
	\begin{tikzpicture}
		\begin{axis}[
			width=\linewidth, 
			height=2.5in,
			xlabel = $t/T_{\mathrm{osc}}$,
			ylabel = $\text{major (minor) axis }\left(r-r_0\right)/r_0$,
			yticklabel style={
				/pgf/number format/fixed,
				/pgf/number format/precision=3
			},
			scaled y ticks=false,
			xtick pos=left, 
			ytick pos=left, 
			xmin = 0,
			xmax =9,
			ymax = 0.09,
			ymin = -0.09,
			legend cell align={left}, 
			legend pos = north east, 
			legend style={draw=none}, 
			cycle list name=color list,
			]
			\addplot[mark size = 1.0pt, color=black, mark=*, mark repeat=1] table [x expr=\thisrowno{0}/0.0222, y expr=\thisrowno{3}+4e-3,col sep=comma]{data/oscillation/csf/eta5e-3_dt1e-5_2.txt};	
			\addplot[mark size = 1.0pt, color=blue, mark=*, mark repeat=1] table [x expr=\thisrowno{0}/0.0222, y expr=\thisrowno{2}+4e-3,col sep=comma]{data/oscillation/csf/eta5e-3_dt1e-5_2.txt};
			
			\addlegendentry{major axis};
			\addlegendentry{minor axis};		
		\end{axis}
              \end{tikzpicture}
              }
	\caption{Damped oscillation of a viscous droplet. The curves show the drop extension along the major and minor axes as functions of time, normalized by the oscillation period, $T_{\mathrm{osc}} = 2\pi/\omega_{2,0} =  2.22\times 10^{-2}\,\si{\second}$.}
	\label{fig:oscillationCSF}
\end{figure}
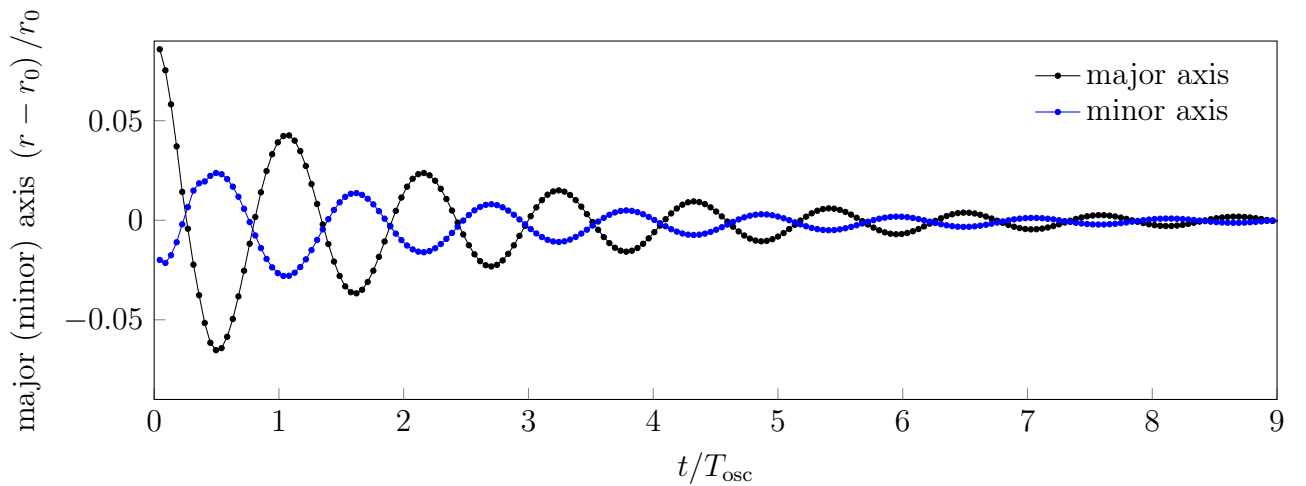

Figure \ref{fig:oscillationCSF} shows the major and minor axis oscillation caused by the initial deformation of the drop from its equilibrium shape, as obtained from the numerical simulation
For a quantitative comparison with the analytical result, we fit the parameters of a damped harmonic oscillator,
\begin{equation}
	r = A_{\mathrm{osc}}\exp\left(-\lambda_{\mathrm{osc}}\right) \cos\left(\omega_{\mathrm{osc}} t\right)\,,
\end{equation}
to the simulation data. We take the amplitude $A_{\mathrm{osc}}$ from the initialization and obtain the best fit for the damping constant $\lambda_{\mathrm{osc}} = 22.44\,\si{\per\second}$  and  the oscillation frequency ${\omega_{\mathrm{osc}} = 261.5\,\si{\per\second}}$. We compare these numerical values with the analytical quantities. For the given material and system parameters, according to Eq.  \eqref{eq:omegaLambda} , they read $\omega_{2,0} = 283,03\,\si{\second}^{-1}$ and $\lambda_2 = 25\,\si{\per\second}$, resulting in the relative deviation $\left|\omega_{2,0} - \omega_{\mathrm{osc}}\right|/ \omega_{2,0}\approx 7\%$ and $\left|\lambda_2-\lambda_\text{osc}\right|/\lambda_2\approx 5.1\%$

An appropriate value of the normal threshold, $\varepsilon^n$, was
found to be crucial for stable oscillations, see Eq.
\eqref{eq:normalFromLiquid}.
Choosing $\varepsilon^\n$ too small results in incorrect curvature computation due to contributions from SPH particles located far away from the interface.  Choosing $\varepsilon^n$ too large results in large statistical errors since only a small number of SPH particles contributes to the  surface tension forces. An appropriate choice of $\varepsilon^n$ compromises between these limits.

\subsection{Test case: Equilibrium shape of a drop in contact with a plane}
\label{sec:sessile}
The numerical simulation of a drop in contact with a plane is a challenging problem. Depending on the choice of the equilibrium contact angle, $\Theta_{\infty}$, which is a parameter of the simulation (see Eq. \eqref{eq:WettingForce}), we describe wetting or de-wetting contact of the liquid drop on the solid substrate. Here, we demonstrate the stability of the numerical method by considering the relaxation of a particular initial state towards equilibrium for varying values of $\Theta_{\infty}$.

The initial condition is described by a hemispherical liquid drop resting on a solid substrate. We place liquid SPH particles on a square lattice with spacing $\Delta x$ inside a hemisphere of radius $r_0 = 10^{-3}\,\si{\metre}$. The flat side of the hemisphere is in contact with the plane, modeled as a circular disk of radius  $3\times 10^{-3}\,\si{\metre}$ and thickness 
$5\Delta x$. In all cases studied, the disk's radius was much larger than the equilibrium radius of the drop.
The disk is represented by solid SPH particles arranged on a square lattice of spacing $\Delta x$.
At time $t=0$, we start the simulation, and the initially hemispherical drop relaxes to its asymptotic equilibrium shape.  For the simulation, we use the parameters in Tab. \eqref{tab:SimParamValid}, the smoothing length $h=2.5\Delta x$, and $\varepsilon^{\mathrm{n}}=0.2/h$. The equilibrium contact angles used to modify the normal vectors are shown in Tab. \ref{table:eqCAsSolidLiquid}. 
\begin{table}[htb]
    \centering 
    \caption{Equilibrium contact angles used to correct the normal vectors of liquid and solid SPH particles in the vicinity of the three-phase contact line. For $\Theta_{\infty}\leq 95^{\circ}$ we employ $\Theta_{\infty}^{\s} \geq \Theta_{\infty}$, see Sec. \ref{section:WettingcontactAngles}}
	\label{table:eqCAsSolidLiquid}
	\begin{tabular}{c | c c c c c c c c c c}
		\toprule
		$\Theta_{\infty}\,/\,^{\circ}$ & 15 & 30 & 45 & 60 & 75 & 90 & 105 & 120 & 135 & 150 \\
		\midrule
		$\Theta_{\infty}^{\s}\,/\,^{\circ}$ & 15 & 32 & 50 & 65 & 85 & 105 & - & - & - & -\\
		\bottomrule
	\end{tabular}
\end{table}

\begin{figure}[htb]
	\centering
	\begin{subfigure}[t]{0.45\linewidth}
          \grafik{
            \includegraphics[width=\linewidth,trim=75 0 75 0,clip]{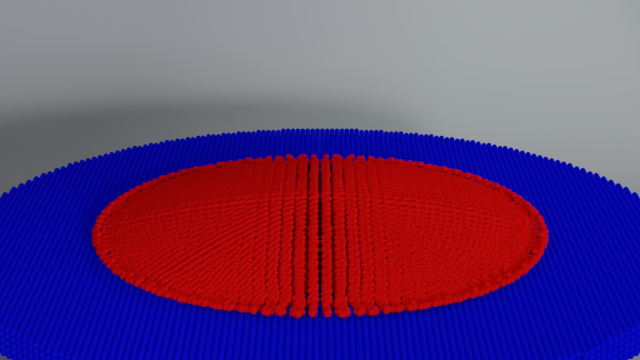}
            }
		\caption{$\Theta_{\infty}^{} = 15^{\circ}$}
	\end{subfigure} 
	\quad
	\begin{subfigure}[t]{0.45\linewidth}
          \grafik{
            \includegraphics[width=\linewidth,trim=75 0 75 0,clip]{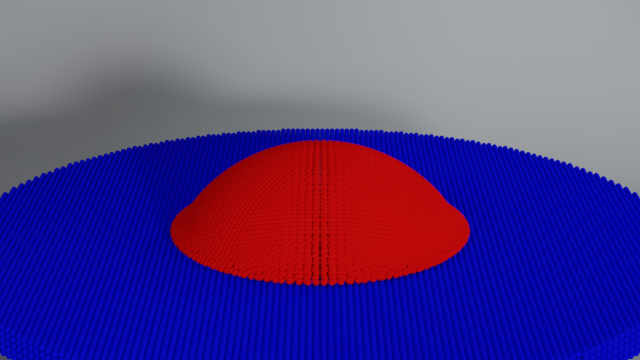}
            }
		\caption{$\Theta_{\infty}^{} = 60^{\circ}$}
	\end{subfigure} 
	\par\medskip
	\begin{subfigure}[t]{0.45\linewidth}
          \grafik{
            \includegraphics[width=\linewidth,trim=75 25 75 0,clip]{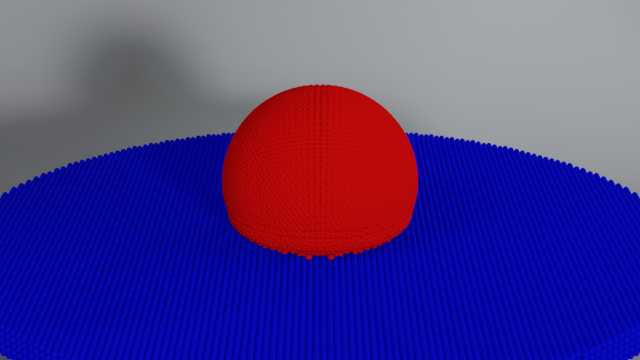}
            }
		\caption{$\Theta_{\infty}^{} = 120^{\circ}$}
	\end{subfigure} 
	\quad
	\begin{subfigure}[t]{0.45\linewidth}
          \grafik{
            \includegraphics[width=\linewidth,trim=75 25 75 0,clip]{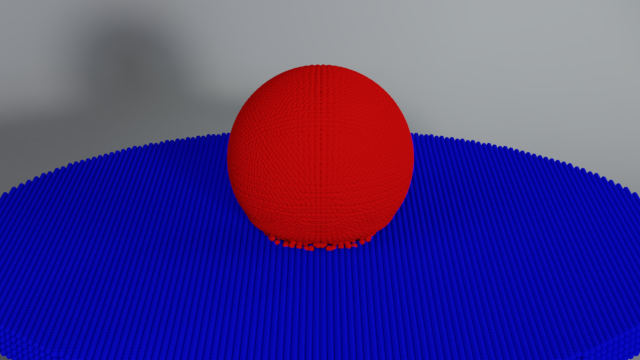}
            }
		\caption{$\Theta_{\infty}^{} = 150^{\circ}$}
	\end{subfigure} 
	\caption{At large time, $t = 1\,\si{\s}$, the drops have assumed their equilibrium shape. Liquid SPH particles are shown in red, and solid particles are shown in blue. For $\Theta_{\infty}^{} \in \{120^{\circ},\ 150^{\circ}\}$, some particles near the three-phase contact line disintegrated from the body of the liquid phase. Refer to Fig. \ref{fig:ContactAngleInterface} for the radial profile of the free surface.}
	\label{fig:equilibriumShapesDrop}
\end{figure}

\begin{figure}[htb]
	\centering 
          \grafik{
	\begin{tikzpicture}
		\begin{axis}[
			width=\linewidth,
			height = 2.5in,
			xlabel = $t\,/\,\si{\second}$,
			ylabel = $\text{kinetic energy}\,/\,\si{\joule}$,
			xtick pos=left, 
			ytick pos=left, 
			legend cell align={left},
			ymode=log,
			xmin = 0.0,
			xmax = 1,
			legend style={draw=none},
			legend columns=2,
			legend style={
				fill=none,
				/tikz/column 2/.style={
					column sep=5pt,
				},
			},
			transpose legend,
			]
			\addplot[ draw=black, thick, mark=none] table [x expr=\thisrowno{0}, y expr=\thisrowno{1},col sep=comma]{data/contactAngle/Ekin_CA15.txt};	
			\addplot[ draw=blue, thick, mark=none] table [x expr=\thisrowno{0}, y expr=\thisrowno{1},col sep=comma]{data/contactAngle/Ekin_CA60.txt};	
			
			\addplot[draw=red, thick, mark=none] table [x expr=\thisrowno{0}, y expr=\thisrowno{1},col sep=comma]{data/contactAngle/Ekin_CA120.txt};	
			\addplot[draw=green, thick, mark=none] table [x expr=\thisrowno{0}, y expr=\thisrowno{1},col sep=comma]{data/contactAngle/Ekin_CA150.txt};

			\addlegendentry{$\Theta_{\infty} = 15^{\circ}$};
			\addlegendentry{$\Theta_{\infty} = 60^{\circ}$};
			\addlegendentry{$\Theta_{\infty} = 120^{\circ}$};
			\addlegendentry{$\Theta_{\infty} = 150^{\circ}$};	
		\end{axis}
              \end{tikzpicture}
              }
	\caption{Total kinetic energy of relaxing droplets as a function of time.  In agreement with physical reality, drops with wetting contact angles relax to a lower value of kinetic energy than drops with non-wetting contact angles. Similarly, the fluctuations are smaller for wetting contact}
	\label{fig:V1Ekin}
\end{figure}
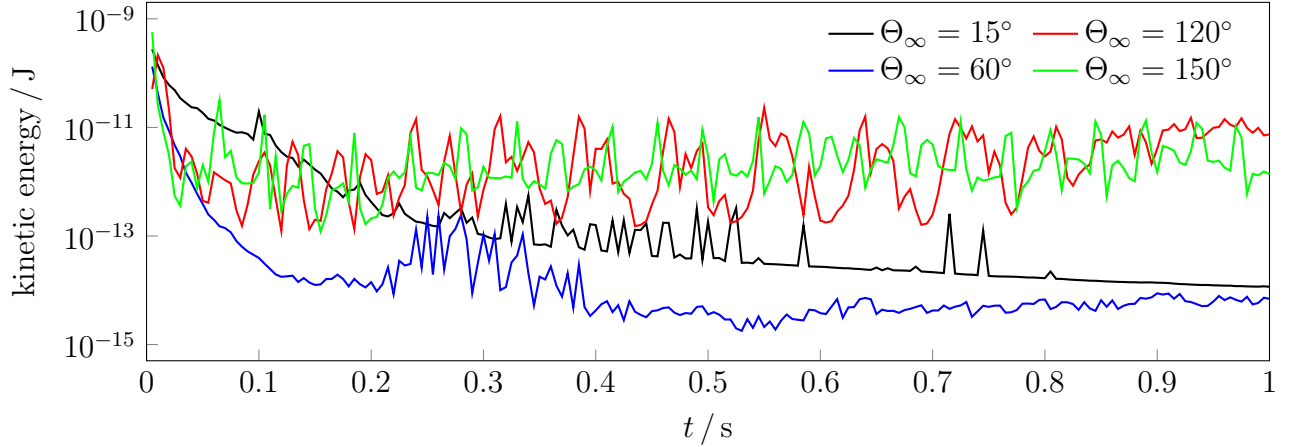

Figure \ref{fig:equilibriumShapesDrop} shows the  drop in equilibrium for contact angles $\Theta_{\infty}^{} \in \{15^{\circ},\ 60^{\circ},\ 120^{\circ},\ 150^{\circ}\}$.
During relaxation, the kinetic energy of the drop and, thus, its constituting SPH particles decays by several orders of magnitude, see Fig. \ref{fig:V1Ekin}.  
We note that the asymptotic value of the kinetic energy of drops with non-wetting exceeds the value for wetting contacts by at least two orders of magnitude: At $t = 1\,\si{\s}$ we find for wetting contact $E_\text{kin}\approx 1.16\times 10^{-14}\,\si{\joule}$ for  $\Theta_{\infty}^{} = 15^{\circ}$ and $E_\text{kin}\approx 6.94\times 10^{-15}\,\si{\joule}$ for   $\Theta_{\infty}^{} = 60^{\circ}$, while for non-wetting contact, we find $E_\text{kin}\approx 7.43\times 10^{-12}\,\si{\joule}$ for $\Theta_{\infty} = 120^{\circ}$ and $E_\text{kin}\approx 1.38\times 10^{-12}\,\si{\joule}$ for $\Theta_{\infty} = 150^{\circ}$. Similarly, the fluctuations of the kinetic energy are much larger for non-wetting contact than for wetting contact. This behavior agrees with physical reality:  Drops of ultrahydrophobic surfaces are much less bound than wetting drops and can reveal interesting dynamics, e.g. the lotus effect \cite{Lafuma:2003}. In terms of the numerical model, this effect can be understood from the influence of the normal threshold, $\varepsilon^n$, used to identify SPH particles with valid normal vectors. The vectors of SPH particles that are located near the three-phase contact line have  magnitude, below the threshold  $\varepsilon^n$ and do, therefore, not experience surface tension. This leads to a vortex flow of particles in this region, which also promotes the translational motion of the liquid drop across the solid substrate which in turn increases the total kinetic energy.

\begin{figure}[htb]
	\centering 
	\begin{subfigure}[b]{0.49\linewidth}
          \grafik{
	\begin{tikzpicture}
		\begin{axis}[
			xlabel = $x/r_0$,
			ylabel = $z/r_0$,
			width=\linewidth,
			xtick pos=left, 
			ytick pos=left, 
			xmin = 0,
			ymin = 0,
			ymax = 2.0,
			xmax = 2.0,
			legend cell align={left}, 
			legend pos = north east, 
			legend style={draw=none,
				fill=none,
				xshift=4pt,
				yshift=4pt}, 
			axis equal image,
			]
			\addplot[draw=black, thick,smooth] 
			table [x expr=\thisrowno{0}/0.001,y expr=\thisrowno{1}/0.001 ,col sep=comma]{data/contactAngle/CA15_interface2.txt};
			
			\addplot[draw=blue, thick] 
			table [x expr=\thisrowno{0}/0.001,y expr=\thisrowno{1}/0.001 ,col sep=comma]{data/contactAngle/CA30_interface2.txt};
			
			\addplot[draw=red, thick] 
			table [x expr=\thisrowno{0}/0.001,y expr=\thisrowno{1}/0.001 ,col sep=comma]{data/contactAngle/CA45_interface2.txt};
			
			\addplot[draw=green, thick] 
			table [x expr=\thisrowno{0}/0.001,y expr=\thisrowno{1}/0.001 ,col sep=comma]{data/contactAngle/CA60_interface2.txt};
			
			\addplot[draw=cyan, thick] 
			table [x expr=\thisrowno{0}/0.001,y expr=\thisrowno{1}/0.001 ,col sep=comma]{data/contactAngle/CA75_interface2.txt};
			
			\addplot[draw=magenta, thick] 
			table [x expr=\thisrowno{0}/0.001,y expr=\thisrowno{1}/0.001 ,col sep=comma]{data/contactAngle/CA90_interface2.txt};
\addlegendentry{$ 16.6^{\circ}$ $(15^{\circ})$};
\addlegendentry{$ 29.8^{\circ}$ $(30^{\circ})$};
\addlegendentry{$46.0^{\circ}$ $(45^{\circ})$};
\addlegendentry{$60.3^{\circ}$ $(60^{\circ})$};
\addlegendentry{$76.6^{\circ}$ $(75^{\circ})$};
\addlegendentry{$91.1^{\circ}$ $(90^{\circ})$};
		\end{axis}
              \end{tikzpicture}
              }
	\caption{Wetting contact angles}
	\label{fig:V1InterfacePositionWetting}
		\end{subfigure}
	\begin{subfigure}[b]{0.49\linewidth}
          \grafik{
	\begin{tikzpicture}
		\begin{axis}[
			xlabel = $x/r_0$,
			ylabel = $z/r_0$,
			xtick pos=left, 
			ytick pos=left, 
			xmin = 0,
			ymin = 0,
			ymax = 2.0,
			xmax = 2.0,
			width=\linewidth,
			legend cell align={left}, 
			legend pos = north east, 
			legend style={draw=none,
				fill=none,
				xshift=4pt,
				yshift=4pt}, 
			axis equal image
			]
			\addplot[draw=blue,thick,shift={(0.0,0.0)}]
			table [x expr=\thisrowno{0}/0.001,y expr=\thisrowno{1}/0.001 ,col sep=comma]{data/contactAngle/CA105_interface3.txt};
			\addplot[draw=red,thick,shift={(0.0,0.0)}]table [x expr=\thisrowno{0}/0.001,y expr=\thisrowno{1}/0.001 ,col sep=comma]{data/contactAngle/CA120_interface3.txt};
			
			\addplot[draw=green,thick]
			table [x expr=\thisrowno{0}/0.001,y expr=\thisrowno{1}/0.001 ,col sep=comma]{data/contactAngle/CA135_interface3.txt};	
			
			\addplot[draw=cyan,thick, smooth]
			table [x expr=\thisrowno{0}/0.001,y expr=\thisrowno{1}/0.001 ,col sep=comma]{data/contactAngle/CA150_interface3.txt};
			
			\addlegendentry{$ 109.5^{\circ}$ $(105^{\circ})$};
			\addlegendentry{$ 122.1^{\circ}$ $(120^{\circ})$};
			\addlegendentry{$ 136.1^{\circ}$ $(135^{\circ})$};	
			\addlegendentry{$ 140.8^{\circ}$ $(150^{\circ})$};
			
		\end{axis}
              \end{tikzpicture}
              }
	\caption{Non-wetting contact angles}
	\label{fig:V1InterfacePositionNonWetting}
	\end{subfigure}
\caption{Cut through a drop  at $y=0$ at large time,  $t=1\,\si{\s}$ , when it assumed its equilibrium shape. The lines show the free liquid-gas interface for different contact angles. The legend shows the momentary value $\Theta$ of the contact angle defined in Eq. \eqref{eq:momentaryTheta} and the equilibrium contact angle, $\Theta_\infty$, in brackets}
\label{fig:ContactAngleInterface}
\end{figure}
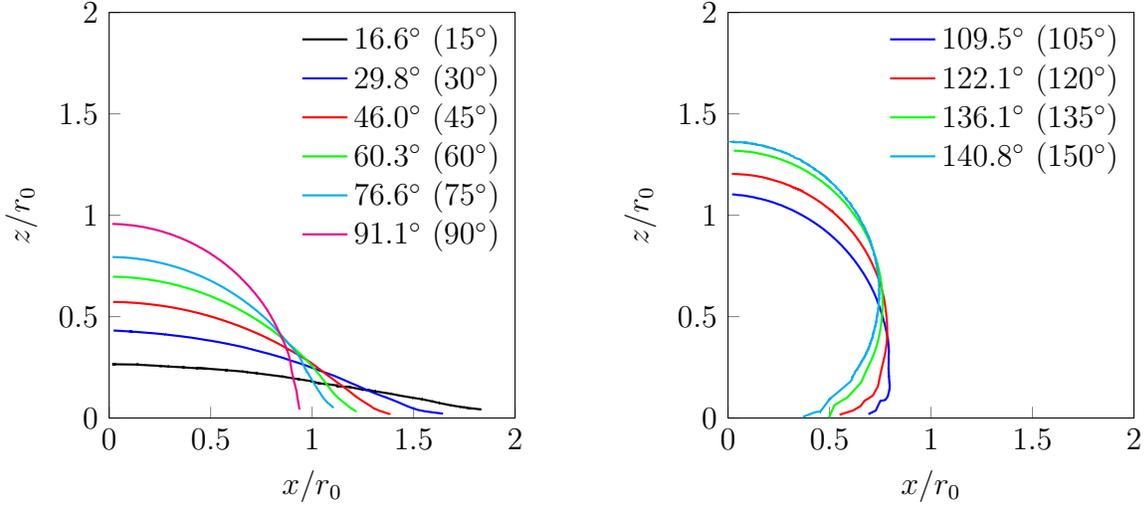

Figure \ref{fig:ContactAngleInterface} shows the position of the free surface of the drop in equilibrium at position $y=0$, that is, a cut through the drop, for wetting and non-wetting contact.  The legend shows in brackets the desired equilibrium contact angle ($\Theta_\infty$)  of the drop (input parameter), and the momentary value
\begin{equation}
		\Theta = \frac{\pi}{2} + \arctan\left(\frac{H-r}{B}\right)
\label{eq:momentaryTheta}
\end{equation}
where $H$ and $B$ are the drop's height and base radius, and 
\begin{equation}
	r = \frac{H^2 + B^2}{2 H}\,
\end{equation}
is the radius of the drop.

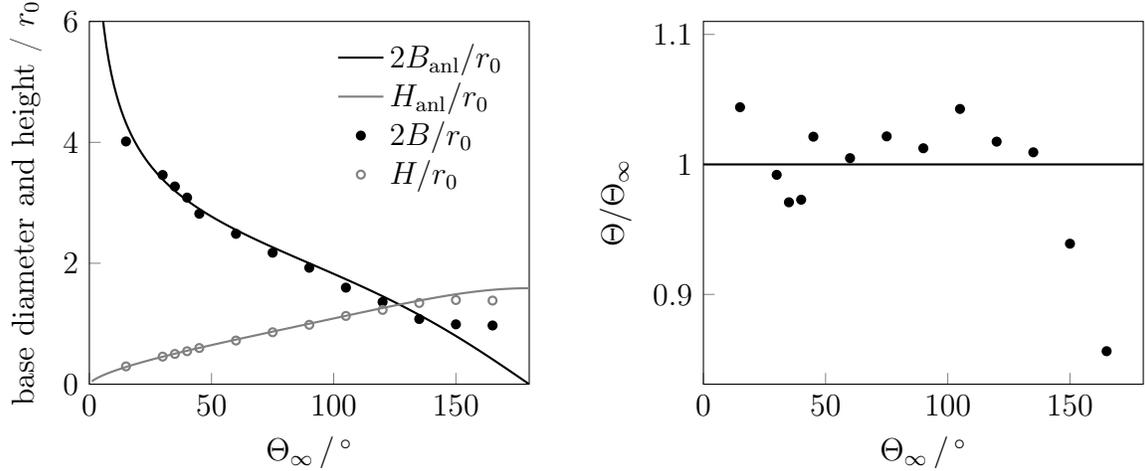
\begin{figure}[htb]
	\centering 
	\begin{subfigure}[t]{0.45\linewidth}
          \grafik{
		\begin{tikzpicture}
			\begin{axis}[
				xlabel = $\Theta_{\infty}\,/\,^{\circ}$,
				ylabel = \text{base diameter and height / $r_0$},
				xtick pos=left, 
				ytick pos=left,
				width = \linewidth,
				xmin = 0,
				xmax = 180,
				ymin = 0,
				ymax = 6,
				legend cell align={left}, 
				legend pos = north east, 
				legend style={draw=none}, 
				]
				\addplot[ color=black,mark=none,thick] 
				table [x index=0, y expr=\thisrowno{4}*2/0.001,col sep=tab]{data/contactAngle/CA_anl.txt}node[pos=0.5,anchor=south, rotate=-20]{};
				
				\addplot[ color=gray,mark=none,thick] 
				table [x index=0, y expr=\thisrowno{3}/0.001 ,col sep=tab, ]{data/contactAngle/CA_anl.txt};
				
				\addplot[ color=black, only marks, mark=*,thick, mark options={scale=0.75}] 
				table [x expr=\thisrowno{0}, y expr=\thisrowno{2}/0.001 ,col sep=comma]{data/contactAngle/widthHeight_csf.txt};
				
				\addplot[color=gray, only marks, mark=o,thick, mark options={scale=0.75}] table [x expr=\thisrowno{0}, y expr=\thisrowno{1}/0.001 ,col sep=comma]{data/contactAngle/widthHeight_csf.txt}node[pos=0.1,anchor=south,rotate=11]{ };
				
				\addlegendentry{$2B_{\anl}/r_0$};
				\addlegendentry{$H_{\anl}/r_0$};
				\addlegendentry{$2B/r_0$};
				\addlegendentry{$H/r_0$};
			\end{axis}
                      \end{tikzpicture}
                      }
		\caption{Analytical (solid lines) and simulated (circles) values of the drop base diameter($2B$) and height ($H$). }
		\label{fig:ComparisonGeometryCA}
	\end{subfigure}\quad
	\begin{subfigure}[t]{0.45\linewidth}
          \grafik{
		\begin{tikzpicture}
			\begin{axis}[
				width = \linewidth,
				xlabel = $\Theta_{\infty}\,/\,^{\circ}$,
				ylabel = $\Theta/\Theta_{\infty}$,
				xtick pos=left, 
				ytick pos=left, 
				xmin = 0,
				xmax = 180,
				ymax = 1.11,
				]
				\addplot[domain=0:180, samples=2,thick]{1};
				\addplot[ color=black,only marks,thick, mark options={scale=0.75}] 
				table [x index=0, y expr=\thisrowno{3}/\thisrowno{0} ,col sep=comma]{data/contactAngle/widthHeight_csf.txt };
				
			\end{axis}
		\end{tikzpicture} 
}		
		\caption{Simulated contact angle, $\Theta$, as a function of the equilibrium contact angle, $\Theta_\infty$}
		\label{fig:GeometryCAError}
	\end{subfigure}
	\caption{Comparison between simulation and theory for the drop's base radius, height, and simulated contact angle, as a function of the equilibrium contact angle}
	\label{fig:AnlCompWH}
\end{figure}

The precision of the numerical model can be evaluated by comparing the simulation results for the drop's height and base radius with the analytical results,
\begin{eqnarray}
	\label{eq:DropBaseRadiusEqAngle}
	B_{\anl} &=& \sqrt{2r_{\anl} H_{\anl} -H_{\anl} ^2}\\
 \label{eq:DropHeightEqAngle}
	H_{\anl} &=& r_{\anl} \left(1-\cos\left(\Theta_{\infty}\right)\right)\,.
\end{eqnarray}
with the analytical value of the drop's radius,
\begin{equation}
	\label{eq:RadiusEqAngle}
	r_{\anl} = \left(\frac{2r_0^3}{2-3\cos(\Theta_{\infty}) + \cos^3\left(\Theta_{\infty}\right)}\right)^{1/3}\,.
\end{equation}
Figure \ref{fig:AnlCompWH} shows this comparison for various contact angle values, $\Theta_\infty$.

In the range ${15^{\circ} \leq\Theta_{\infty} \leq 135^{\circ}}$, the deviation of the simulated contact angle from the analytical value is below  $4.5\,\%$, see Fig. \ref{fig:GeometryCAError}, and increases to $14.3\,\%$ for ${\Theta_{\infty} \leq 150^{\circ}}$. This deviation is due to the unprecise contributions to the force from SPH particles located near the three-phase contact line. The precision could be improved by decreasing $\varepsilon^{\n}$, however, this would deteriorate the accuracy of the computed curvature. As in the preceding example, we have to compromise between large and small values of $\varepsilon^{\n}$.

\subsection{Test case: Droplet deformation due to gravity}
\label{sec:case_gravity}
In the test cases presented so far, we did not consider the action of gravity. In the current test case, we study the deformation of a drop resting on a horizontal plate as a function of the value of the gravity constant, $g$.  Gravity gives rise to a body-force acceleration, ${\vec{f}^g = \left[0,\ 0,\ -g\right]}$, acting on the SPH particles. This force leads to flattening the drop, which shall be studied here.

We use the properties given in Tab. \ref{tab:SimParamValid} and the setup geometry introduced in Sec. \ref{sec:sessile}:  A total of 16\,776 liquid SPH particles are placed on a square lattice inside a hemisphere of radius $r_0 = 10^{-3}\,\si{\metre}$, resting on a plane solid substrate of radius $4\times 10^{-3}\,\si{\metre}$. The equilibrium contact angle is $\Theta_{\infty} = 50^{\circ}$ ($\Theta_{\infty}^{\s} = 55^{\circ}$).

\begin{figure}[htb]
	\centering 
          \grafik{
	\begin{tikzpicture}
		\begin{axis}[
			width = 0.6\linewidth,
			xlabel = $\mathrm{Bo}$,
			ylabel = $H/H_0$,
			xtick pos=left, 
			ytick pos=left, 
			xmode=log,
			xmin = 0.001,
			ymax = 1.2,
			xmax = 100,
			legend cell align={left}, 
			legend pos = south west, 
			legend style={draw=none}, 
			]
			\addplot[color=black, thick,forget plot, mark options={scale=0.75}] 
			table [x expr=\thisrowno{0},y expr=\thisrowno{1}/0.000631
			,col sep=tab]{data/EoGrav2/EoGrav_anl_50_130_2.txt}node[pos=0.94,anchor=south,rotate=-62,yshift=-15pt]{$H_{\infty}/H_0$};
			
			\addplot[thick,domain=0.001:100, samples=2,thick]{1}node[pos=0.85,anchor=south,rotate=0,yshift=-3pt]{$H_0/H_0$};
			\addplot[color=blue, only marks, thick, mark options={scale=0.75}] 
			table [x expr=\thisrowno{0},y expr=\thisrowno{1}/0.000631 ,col sep=space]{data/EoGrav2/EoGrav_sph_50_130.txt};

		\end{axis}
              \end{tikzpicture}
              }
	\caption{Drop height as a function of the Bond number. For small gravity, $\Bo\to 0$, the regime is dominated by surface tension, and  $H$ approaches $H_0$. For large gravity, $\Bo\to \infty$, the regime is gravity-dominated,  and $H$ approaches $H_\infty$ given by Eq. \eqref{eq:Hinfty}. The symbols show the drop height obtained from SPH simulations, and the solid lines show the limits $H_0$ and $H_\infty$.}
	\label{fig:dropGravitya}
\end{figure}
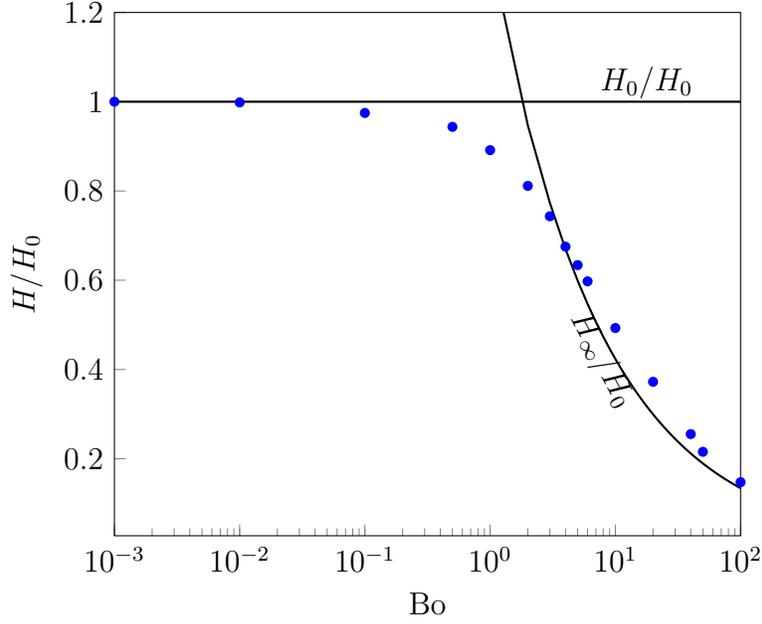

Figure \ref{fig:dropGravitya} shows the height of the drop, $H$, as a function of the acceleration due to gravity, $g$.  Here, gravity is expressed  by the Bond number that quantifies the ratio of gravitational to surface tension forces:
\begin{equation}
	\Bo \equiv \frac{\rho g r_0^2}{\sigma}\,.
\end{equation}
The drop height, $H$, is scaled by the drop height in the absence of gravity $H_0$ where $\Bo =0$.
In Fig. \ref{fig:dropGravitya}, the Bond number covers the interval $\Bo\in[10^{-3},10^2]$, and the function $H(\Bo)$ decays monotonously. For small $\Bo$, the situation is surface-tension dominated, thus, $H\to H_0$ for $\Bo\to 0$. 

\begin{figure}[htb]
	\centering
          \grafik{
	\begin{tikzpicture}
		\begin{axis}[
			width = 0.6\linewidth,
			xlabel = $x/H_0$,
			ylabel = $z/H_0$,
			xtick pos=left, 
			ytick pos=left, 
			xmin = 0,
			ymin = 0,
			ymax = 2.0,
			xmax = 4.0,
			legend cell align={left}, 
			legend pos = north east, 
			legend style={draw=none}, 
			axis equal image,
			]
			\addplot+[draw=black, thick, mark=none] 
			table [x expr=\thisrowno{0}/0.00062010,y expr=\thisrowno{1}/0.00062010 ,col sep=comma]{data/EoGrav2/interface/Bo_1e-3_interface.txt};
			\addplot+[draw=blue, thick, mark=none] 
			table [x expr=\thisrowno{0}/0.00062010,y expr=\thisrowno{1}/0.00062010 ,col sep=comma]{data/EoGrav2/interface/Bo_1_interface.txt};
			
			\addplot+[draw=red, thick, mark=none] 
			table [x expr=\thisrowno{0}/0.00062010,y expr=\thisrowno{1}/0.00062010 ,col sep=comma]{data/EoGrav2/interface/Bo_5_interface.txt};
			
			\addplot+[draw=green, thick, mark=none] 
			table [x expr=\thisrowno{0}/0.00062010,y expr=\thisrowno{1}/0.00062010 ,col sep=comma]{data/EoGrav2/interface/Bo_40_interface.txt};
			\addlegendentry{$\mathrm{Bo} = 10^{-3}$};
			\addlegendentry{$\mathrm{Bo} = 1$};
			\addlegendentry{$\mathrm{Bo} = 5$};
			\addlegendentry{$\mathrm{Bo}= 40$};	
		\end{axis}
              \end{tikzpicture}
              }
	\caption{Cut through the drop at $y=0$ (liquid-gas interface position) as a function of the Bond number for $\Theta_{\infty}= 50^{\circ}$.}	
 \label{fig:InterfacialParticlesGravity}
\end{figure}
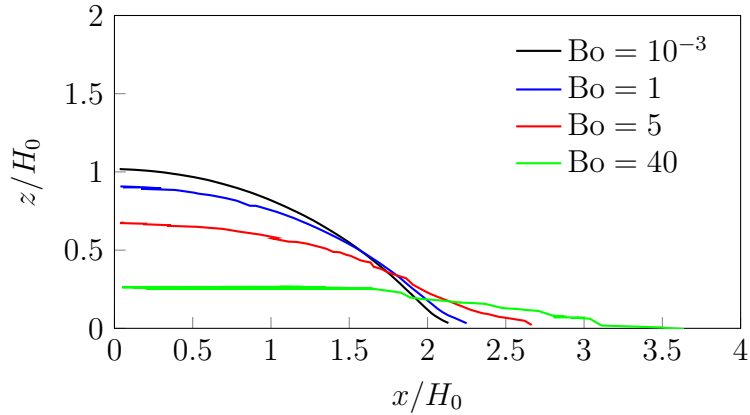

For large gravity, $\Bo\to \infty$, the regime is gravity-dominated, and he height of the drop converges to the capillary length \cite{Dupont2010}
\begin{equation}
	H_\infty = 2 \sqrt{\frac{\sigma}{\rho g_z}} \sin{\left(\frac{\Theta_{\infty}}{2}\right)}\,.
 \label{eq:Hinfty}
\end{equation}
Figure \ref{fig:InterfacialParticlesGravity} shows the drop profile for several values of $\Bo$, that is, the position of the free surface at $y=0$. For real-life applications, a dynamic contact angle model (such as \cite{GOHL2018}) may be implemented for taking into account chemical heterogeneity of the substrate. 

\subsection{Test case: Droplet pinning}

\begin{figure}[htb]
	\centering
          \grafik{
	\begin{tikzpicture}
		\begin{axis}[
			width=\linewidth,
			scale only axis,
			xmin=0,
			xmax=675.25,
			ymin=0,
			ymax=205,
			axis equal image,
			axis line style={draw=none},
			tick style={draw=none},
			xticklabels={,,},
			yticklabels={,,},
			inner sep=0em,
			]
			\addplot[thick,blue] graphics[xmin=0,ymin=0,xmax=675.25,ymax=200] {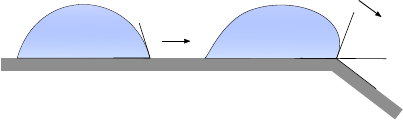};
			\draw[latex-latex] (195,102) to[out=90,in=-155] (235,155);
			\node[text width = 0.5cm] at (225,118) {$\Theta_{\infty}$};
			
			\draw[latex-latex] (520,102) to[out=90,in=-200] (581,145);
			\node[text width = 0.5cm] at (560,122) {$\Theta<\Theta_\infty+\Psi$};	
			
			\draw[latex-latex] (630,50) to[out=60,in=-90] (640,100);
			\node[text width = 0.5cm] at (615,85) {$\Psi$};		
			\node[text width=0.5cm,anchor=west] at (620,192){$\vec{f}^{\mathrm{b}} $};
		\end{axis}
              \end{tikzpicture}
              }
	\caption{Sketch a pinned drop at the contact line between two planes of different inclinations. The applied body acceleration $\vec{f}^{\mathrm{b}}$ drives the droplet toward the contact line between the planes. It can pass the barrier only if  the contact angle exceeds the threshold $\Theta = \Theta_{\infty} + \Psi$}	
	\label{fig:sketchPinning}
\end{figure}

The contact line of  two planes with different inclinations represents a barrier for liquid droplets. To pass the barrier, the droplet's contact angle must exceed $\Theta_\infty + \Psi$ \cite{Breinlinger2013}, where $\Psi$ is the difference between the inclinations of the planes. To overcome this threshold, sufficient force is needed. In the absence of such a force, the droplet remains attached at the contact line of the planes, see Fig. \ref{fig:sketchPinning}.

We describe the numerical experiment similar to Secs. \ref{sec:sessile} and \ref{sec:case_gravity} with the parameters specified in Tab. \ref{tab:SimParamValid}, viscosity $\eta = 5\times 10^{-3}\,\si{\pascal\second}$, and the equilibrium contact angle $\Theta_{\infty} = 75^{\circ}$ ($\Theta_{\infty}^{\s} = 80^{\circ}$).

Similar to the preceding test cases, we fill a hemisphere with radius $r_0=10^{-3}\,\si{\metre}$ with 16\,776 liquid SPH particles arranged on a square lattice. This half-sphere rests on a solid horizontal plane modeled as a block of size $(2.6\times3 \times0.25)\,\si{\cubic\milli\metre}$ filled by SPH particles arranged in a square lattice. The initial condition is sketched in the left part of Fig. \ref{fig:sketchPinning}. The second plane is modeled in the same way but inclined by $\Psi \in \{22.5^{\circ},\ 45^{\circ},\ 67.5^{\circ}\}$, see right part of Fig. \ref{fig:sketchPinning}. 

\begin{figure}[htb]
	\centering
	\quad \begin{subfigure}[t]{0.3\linewidth}
          \grafik{
		\begin{overpic}[width=\linewidth,trim=50 0 50 50,clip]{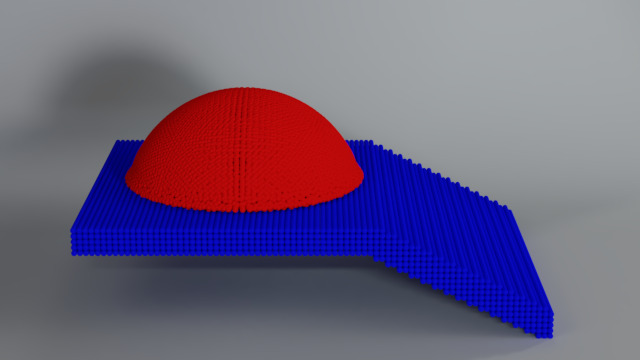}
			\put(-10,5){\rotatebox{90}{$\Psi = 22.5^{\circ}$}}
                      \end{overpic}
                      }
	\end{subfigure}
	\begin{subfigure}[t]{0.3\linewidth}
             \grafik{
		\includegraphics[width=\linewidth,trim=50 0 50 50,clip]{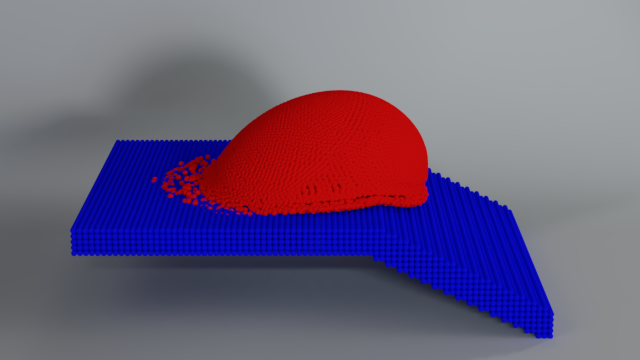}
                }
              \end{subfigure}
	\begin{subfigure}[t]{0.3\linewidth}
          \grafik{
		\includegraphics[width=\linewidth,trim=50 0 50 50,clip]{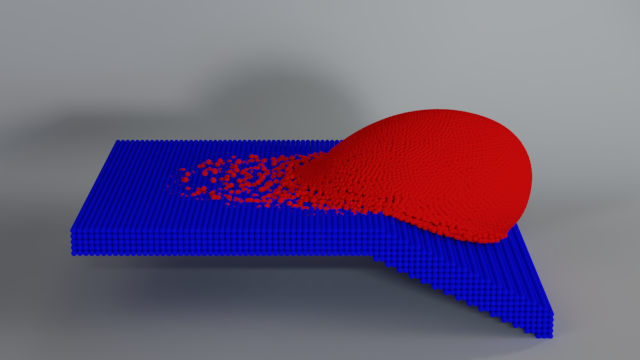}
                }
	\end{subfigure}
	
	\quad	\begin{subfigure}[t]{0.3\linewidth}
          \grafik{
		\begin{overpic}[width=\linewidth,trim=50 0 50 50,clip]{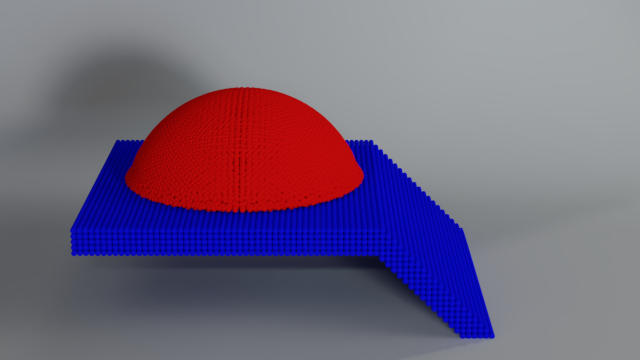}
			\put(-10,5){\rotatebox{90}{$\Psi = 45^{\circ}$}}
                      \end{overpic}
                      }
	\end{subfigure}
	\begin{subfigure}[t]{0.3\linewidth}
          \grafik{
		\includegraphics[width=\linewidth,trim=50 0 50 50,clip]{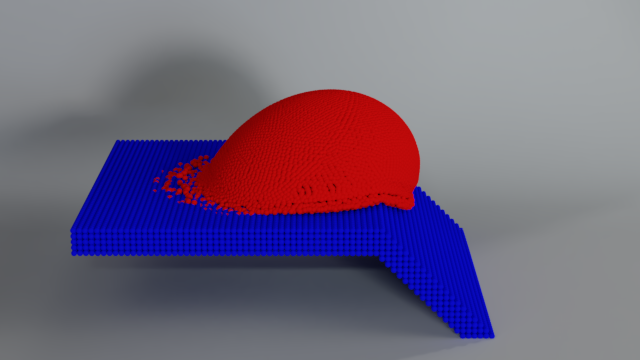}
                }
	\end{subfigure}
	\begin{subfigure}[t]{0.3\linewidth}
          \grafik{
		\includegraphics[width=\linewidth,trim=50 0 50 50,clip]{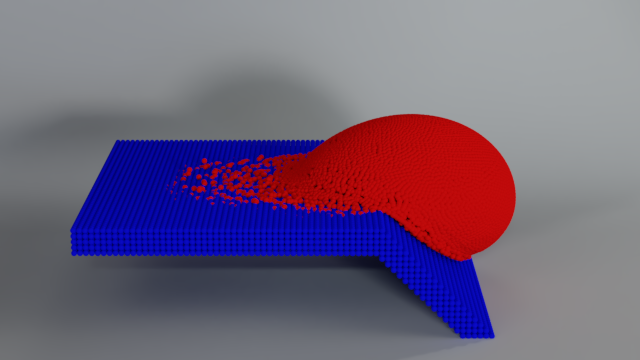}
                }
	\end{subfigure}
	
	\quad 	\begin{subfigure}[t]{0.3\linewidth}
          \grafik{
		\begin{overpic}[width=\linewidth,trim=50 0 50 50,clip]{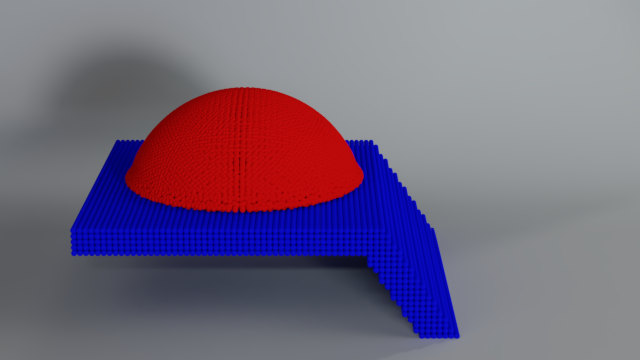}
			\put(-10,5){\rotatebox{90}{$\Psi = 67.5^{\circ}$}}
                      \end{overpic}
                      }
	\end{subfigure}
	\begin{subfigure}[t]{0.3\linewidth}
          \grafik{
		\includegraphics[width=\linewidth,trim=50 0 50 50,clip]{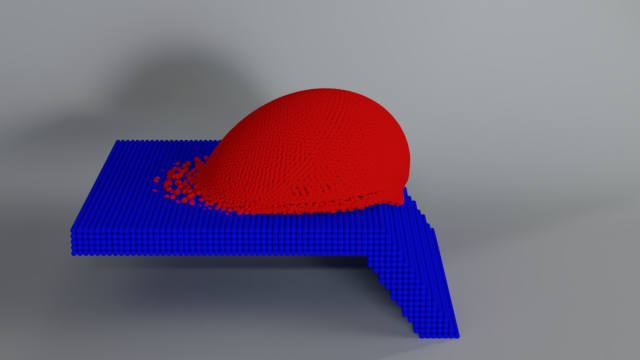}
                }
	\end{subfigure}
	\begin{subfigure}[t]{0.3\linewidth}
          \grafik{
		\includegraphics[width=\linewidth,trim=50 0 50 50,clip]{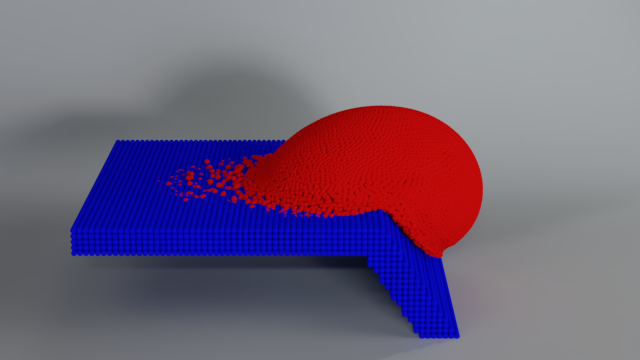}
                }
	\end{subfigure}
	
	\quad 	\begin{subfigure}[t]{0.3\linewidth}
          \grafik{
		\begin{overpic}[width=\linewidth,trim=50 0 50 50,clip]{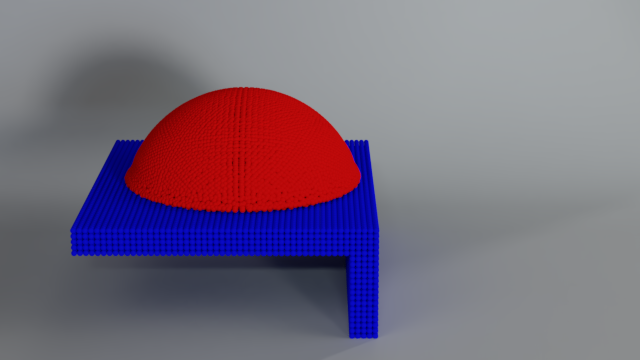}
			\put(-10,5){\rotatebox{90}{$\Psi = 90^{\circ}$}}
                      \end{overpic}
                      }
		\caption{$t = 50\,\si{\milli\second}$}
	\end{subfigure}
	\begin{subfigure}[t]{0.3\linewidth}
          \grafik{
		\includegraphics[width=\linewidth,trim=50 0 50 50,clip]{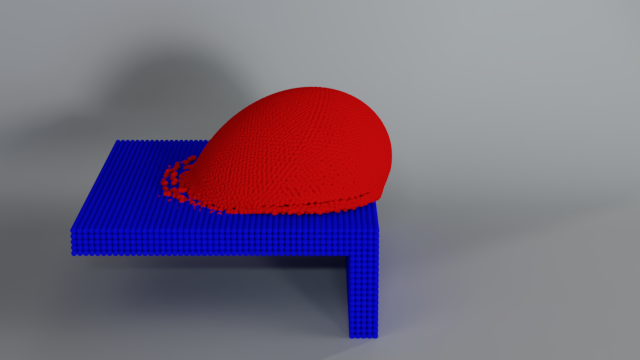}
		\caption{$t = 100\,\si{\milli\second}$}
                }
	\end{subfigure}
	\begin{subfigure}[t]{0.3\linewidth}
             \grafik{
		\includegraphics[width=\linewidth,trim=50 0 50 50,clip]{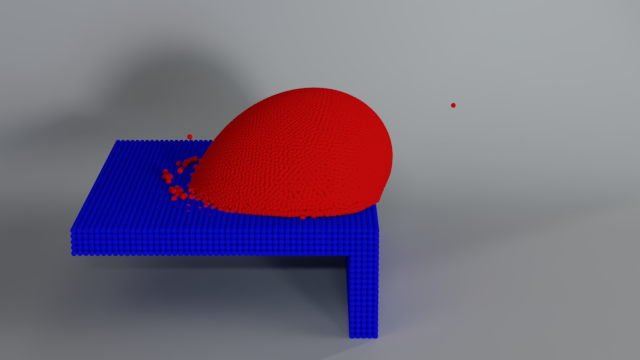}
		\caption{$t = 150\,\si{\milli\second}$}
                }
	\end{subfigure}
	\caption{Simulation snapshots of droplets that are accelerated towards the contact line of two planes with different inclinations. Column (a) shows the droplets at time $t = 50\,\si{\milli\second}$ resting in equilibrium on a horizontal solid plane with the equilibrium contact angle $\Theta_{\infty} = 50^{\circ}$. Column (b) shows the droplets at time $t = 100\,\si{\milli\second}$ when approaching the contact line under the action of acceleration $\vec{f}^{\mathrm{b}}$. Column (c) at time $t = 150\,\si{\milli\second}$: For $\Psi\in\{22.5^\circ, 45^\circ, 67.5^\circ\}$ the contact angle exceeds the threshold, thus, the drop passed over the surface discontinuity. For $\Psi = 90^{\circ}$, the contact angle is below the threshold, thus, the drop remains pinned}
	\label{fig:pinning}
\end{figure}

To prepare our initial conditions, we simulate relaxing the drop on the horizontal substrate to equilibrium.  After  $t = 50\,\si{\milli\second}$ of relaxation, the equilibrium shape was assumed.
For   $t > 50\,\si{\milli\second}$ a constant body acceleration, $\vec{f}^{\mathrm{b}} = [7.5,\ 0,\ -7.5]$, drives the drop towards the contact line of the planes. Figure \ref{fig:pinning} shows snapshots of the simulation. 

Column (a) shows the equilibrated drops at $t = 50\,\si{\milli\second}$.  Column (b) shows the drops at $t=100\,\si{\milli\second}$ under the influence of the acceleration $\vec{f}^{\mathrm{b}}$. Due to the weighting concept of the SPH method, pinning starts to occur when the contact line of the two planes enters the smoothing length of an SPH particle. Instead of a sharp force, the particles experience a smooth counteracting force, enabling some SPH particles to pass the contact line. The last column of Fig. \ref{fig:pinning} shows the droplets at $t = 150\,\si{\milli\second}$. For $\Psi \in \{22.5^{\circ},\ 45^{\circ},\ 67.5^{\circ}\}$ the drop passes the discontinuity. For $\Psi = 90$, the external acceleration is insufficient to generate the threshold contact angle  $\Theta_{\infty} + \Psi = 165^{\circ}$.

\begin{figure}[htb]
	\centering
  \grafik{
	\begin{tikzpicture}
		\begin{axis}[
			width=0.8\linewidth,
			height = 2in,
			xlabel = $t\,/\,\si{\milli\second}$,
			ylabel = kinetic energy$\,/\,\si{\joule}$,
			xtick pos=left, 
			ytick pos=left, 
			legend cell align={left},
			ymode=log,
			xmax=150,
			xmin = 0.0,
			ymax = 1e-7,
			legend style={draw=none},
			legend pos = north west,
			legend columns=2,
			legend style={
				fill=none,
				/tikz/column 2/.style={
					column sep=5pt,
				},
			},
			transpose legend,
			]
			\addplot[ draw=black, thick, mark=none] table [x expr=\thisrowno{0}*1000, y expr=\thisrowno{1},col sep=comma]{images/pinning_csf/psi22.5/Ekin.txt};	
			\addplot[ draw=blue, thick, mark=none] table [x expr=\thisrowno{0}*1000, y expr=\thisrowno{1},col sep=comma]{images/pinning_csf/psi45.0/Ekin.txt};	
			\addplot[ draw=red, thick, mark=none] table [x expr=\thisrowno{0}*1000, y expr=\thisrowno{1},col sep=comma]{images/pinning_csf/psi67.5/Ekin.txt};	
			\addplot[ draw=green, thick] table [x expr=\thisrowno{0}*1000, y expr=\thisrowno{1},col sep=comma]{images/pinning_csf/psi90.0/Ekin.txt};	
			\addlegendentry{$\Psi = 22.5^{\circ}$};
			\addlegendentry{$\Psi = 45^{\circ}$};
			\addlegendentry{$\Psi = 67.5^{\circ}$};
			\addlegendentry{$\Psi = 90^{\circ}$};
		\end{axis}
              \end{tikzpicture}
              }
	\caption{Kinetic energies of the drops over time for different inclinations, $\Psi$, of the plane. The time $t$< 50\,\si{\milli\second} corresponds to the relaxation of the drop to find its equilibrium shape.  For $\Psi\in\{22.5^{\circ},\ 45^{\circ},\ 67.5^{\circ}\}$ the drop passes the discontinuity. For $\Psi = 90$ it remains pinned}
	\label{fig:ContactDiscontinuityEkin}
\end{figure}
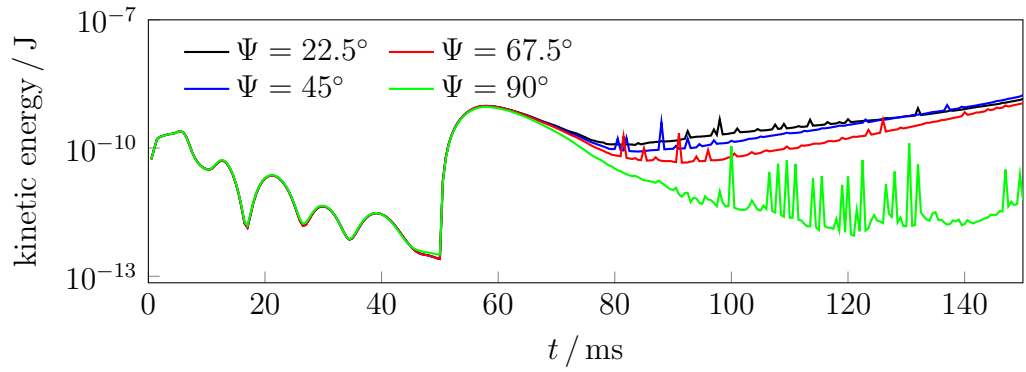

Figure \ref{fig:ContactDiscontinuityEkin} shows the total kinetic energy of the liquid SPH particles as a function of time. 
During the relaxation, $t\in [0,50\,\si{\milli\second}]$,  the drop finds its equilibrium shape, therefore, energy decays. For $t>50\,\si{\milli\second}$, the decline is accelerated by  $\vec{f}^{\mathrm{b}}$ acting on the liquid SPH particles leading to a sharp increase in the total kinetic energy.  At $t\approx 55\,\si{\milli\second}$, the drop approaches the contact line between the two planes, indicated by a decrease of the kinetic energy for $t > 55\,\si{\milli\second}$.  As expected, the smallest deceleration of the drop is experienced for  $\Psi =22.5\,^{\circ}$ and the most significant for $\Psi = 90\,^{\circ}$. The subsequent increase in kinetic energy  for $t \gtrsim 85\,\si{\milli\second}$,  for $\psi<90\,^{\circ}$ corresponds to the drop's passing of the discontinuity. For $\Psi = 90\,^{\circ}$, the kinetic energy remains about three orders of magnitude below, indicating the pinning of the drop at the contact line.

The kinetic energy curves in Fig.  \ref{fig:ContactDiscontinuityEkin} show sharp peaks. These peaks are caused by SPH particles on the rear side of the moving droplet. Due to the small number of liquid SPH particles attached to the solid phase behind the moving droplet, the calculated curvatures of these particles can adopt large, inaccurate values. As a result, these particles experience either high accelerations toward the solid substrate or into the surrounding environment. The wall boundary condition prevents liquid SPH particles from entering the solid plane due to the large pressure of the solid SPH particles. However, this also leads to a strong repulsion of the liquid particles into the environment. This problem could be solved by calculating the mean curvature of the liquid SPH particles only when there is a sufficient number of neighbors of the liquid phase. Otherwise, the particles should remain unaffected by the surface tension.

\section{Conclusion}
We introduce a surface tension model based on the CSF approach and applicable to three-dimensional free surface flows.  In contact with a substrate, the model can handle wetting and non-wetting behavior in a wide range of parameters. This is made possible with the use of primarily 3 empirical parameters, namely the Shephard sum correction for free surface introduced in Eq. 37, the threshold for magnitude of surface normal vectors (Eq. 31), $\epsilon^\textrm{n}$ and the desired equilibrium contact angle associated with the substrate $\Theta_\infty^s$ listed in Table 2.  We have shown that the model performs numerically stable in several test cases. By directly comparing analytical results and literature results, we could show that the model delivers quantitatively reliable results.

To demonstrate the model's performance, we applied the new simulation method to a set of free surface problems, some of which are notoriously difficult to simulate. These test examples include (a) plane Poiseuille flow, (b) the Laplace jump of a droplet, (c) the oscillation and relaxation of a perturbed droplet, (d) the equilibrium shape of a droplet in contact with a wetting/non-wetting substrate and its relaxation to this equilibrium, (e) the flattening of a droplet under the action of gravity, and (f) the interaction of a droplet with a barrier under the action of gravity. 

For all of the test cases, we provide detailed quantitative comparisons with analytical results and results from the literature and discuss the reasons for deviations.

\section*{Acknowledgment}
This work was funded by the Deutsche Forschungsgemeinschaft (DFG, German Research Foundation)-Project-ID 61375930-SFB 814 ``Additive Manufacturing'' TP B1. We thank the Gauss Centre for Supercomputing for providing computer time through the John von Neumann Institute for Computing on the GCS Supercomputer JUWELS at J\"ulich Supercomputing Centre. The authors gratefully acknowledge the scientific support and HPC resources provided by the Erlangen National High-Performance Computing Center (NHRFAU) of the Friedrich-Alexander-Universität Erlangen-Nürnberg (FAU). The hardware is funded by the German Research Foundation (DFG). The work was also supported by the Interdisciplinary Center for Nanostructured Films (IZNF), the Competence Unit for Scientific Computing (CSC), and the Interdisciplinary Center for Functional Particle Systems (FPS) at Friedrich-Alexander University Erlangen-N\"urnberg.  The work was also partially supported by the Science Education and Research Board (SERB) through the Startup Research Grant (SRG) number SRG\/2022\/000436.

\bibliographystyle{elsarticle-num}
\bibliography{surface-tension-SPH}

\begin{thebibliography}{10}
\expandafter\ifx\csname url\endcsname\relax
  \def\url#1{\texttt{#1}}\fi
\expandafter\ifx\csname urlprefix\endcsname\relax\def\urlprefix{URL }\fi
\expandafter\ifx\csname href\endcsname\relax
  \def\href#1#2{#2} \def\path#1{#1}\fi

\bibitem{Babadagli2002}
T.~Babadagli, Dynamics of capillary imbibition when surfactant, polymer, and hot water are used as aqueous phase for oil recovery, J. Colloid Interface Sci. 246 (2002) 203--213.
\newblock \href {https://doi.org/10.1006/jcis.2001.8015} {\path{doi:10.1006/jcis.2001.8015}}.

\bibitem{Babadagli2005}
T.~Babadagli, Y.~Boluk, Oil recovery performances of surfactant solutions by capillary imbibition, J. Colloid Interface Sci. 282 (2005) 162--175.
\newblock \href {https://doi.org/10.1016/j.jcis.2004.08.149} {\path{doi:10.1016/j.jcis.2004.08.149}}.

\bibitem{Ebadian2011}
M.~A. Ebadian, C.~X. Lin, A review of high-heat-flux heat removal technologies, J. Heat Transfer 133, 110801 (2011).
\newblock \href {https://doi.org/10.1115/1.4004340} {\path{doi:10.1115/1.4004340}}.

\bibitem{Calvert2001}
P.~Calvert, Inkjet printing for materials and devices, Chem. Mater. 13 (2001) 3299--3305.
\newblock \href {https://doi.org/10.1021/cm0101632} {\path{doi:10.1021/cm0101632}}.

\bibitem{Kam_2012}
D.~H. Kam, S.~Bhattacharya, J.~Mazumder, Control of the wetting properties of an {AISI} 316l stainless steel surface by femtosecond laser-induced surface modification, J. Micromech. Microeng. 22 (2012) 105019.
\newblock \href {https://doi.org/10.1088/0960-1317/22/10/105019} {\path{doi:10.1088/0960-1317/22/10/105019}}.

\bibitem{Lai2013}
Y.~Lai, F.~Pan, C.~Xu, H.~Fuchs, L.~Chi, In situ surface-modification-induced superhydrophobic patterns with reversible wettability and adhesion, Adv. Materials 25 (2013) 1682--1686.
\newblock \href {https://doi.org/10.1002/adma.201203797} {\path{doi:10.1002/adma.201203797}}.

\bibitem{Lucy1977}
L.~B. Lucy, A numerical approach to the testing of the fission hypothesis, Astron. J. 82 (1977) 1013.
\newblock \href {https://doi.org/10.1086/112164} {\path{doi:10.1086/112164}}.

\bibitem{GingoldMonaghan1977}
R.~A. Gingold, J.~J. Monaghan, Smoothed particle hydrodynamics: {T}heory and application to non-spherical stars, Mon. Not. R. Astron. Soc. 181 (1977) 375--389.
\newblock \href {https://doi.org/10.1093/mnras/181.3.375} {\path{doi:10.1093/mnras/181.3.375}}.

\bibitem{Nugent2000}
S.~Nugent, H.~A. Posch, Liquid drops and surface tension with smoothed particle applied mechanics, Phys. Rev. E 62 (2000) 4968.
\newblock \href {https://doi.org/10.1103/physreve.62.4968} {\path{doi:10.1103/physreve.62.4968}}.

\bibitem{Tartakovsky2005}
A.~Tartakovsky, P.~Meakin, Modeling of surface tension and contact angles with smoothed particle hydrodynamics, Phys. Rev. E 72 (2005) 026301.
\newblock \href {https://doi.org/10.1103/PhysRevE.72.026301} {\path{doi:10.1103/PhysRevE.72.026301}}.

\bibitem{Tartakovsky2016}
A.~M. Tartakovsky, A.~Panchenko, Pairwise force smoothed particle hydrodynamics model for multiphase flow: {S}urface tension and contact line dynamics, J. Comput. Phys. 305 (2016) 1119--1146.
\newblock \href {https://doi.org/10.1016/j.jcp.2015.08.037} {\path{doi:10.1016/j.jcp.2015.08.037}}.

\bibitem{nair2018}
P.~Nair, T.~P{\"o}schel, Dynamic capillary phenomena using incompressible {SPH}, Chem. Eng. Sci. 176 (2018) 192--204.
\newblock \href {https://doi.org/10.1016/j.ces.2017.10.042} {\path{doi:10.1016/j.ces.2017.10.042}}.

\bibitem{Brackbill1992}
J.~U. Brackbill, D.~B. Kothe, C.~Zemach, A continuum method for modeling surface tension, J. Comput. Phys. 100 (1992) 335--354.
\newblock \href {https://doi.org/10.1016/0021-9991(92)90240-Y} {\path{doi:10.1016/0021-9991(92)90240-Y}}.

\bibitem{Morris2000}
J.~P. Morris, Simulating surface tension with {S}moothed {P}article {H}ydrodynamics, Int. J. Numer. Methods Fluids 33 (2000) 333--353.
\newblock \href {https://doi.org/10.1002/1097-0363(20000615)33:3<333::AID-FLD11>3.0.CO;2-7} {\path{doi:10.1002/1097-0363(20000615)33:3<333::AID-FLD11>3.0.CO;2-7}}.

\bibitem{SIROTKIN2012}
F.~V. Sirotkin, J.~J. Yoh, A new particle method for simulating breakup of liquid jets, Journal of Computational Physics 231 (2012) 1650--1674.
\newblock \href {https://doi.org/10.1016/j.jcp.2011.10.020} {\path{doi:10.1016/j.jcp.2011.10.020}}.

\bibitem{Rowlinson2013}
J.~S. Rowlinson, B.~Widom, Molecular Theory of Capillarity, Courier Corporation, 2013.
\newblock \href {https://doi.org/10.1002/bbpc.19840880621} {\path{doi:10.1002/bbpc.19840880621}}.

\bibitem{shigorina2017smoothed}
E.~Shigorina, J.~Kordilla, A.~M. Tartakovsky, Smoothed particle hydrodynamics study of the roughness effect on contact angle and droplet flow, Phys. Rev. E 96 (2017) 033115.
\newblock \href {https://doi.org/10.1103/PhysRevE.96.033115} {\path{doi:10.1103/PhysRevE.96.033115}}.

\bibitem{bao2019modified}
Y.~Bao, L.~Li, L.~Shen, C.~Lei, Y.~Gan, Modified smoothed particle hydrodynamics approach for modelling dynamic contact angle hysteresis, Acta Mech. Sin. 35 (2019) 472--485.
\newblock \href {https://doi.org/10.1007/s10409-018-00837-8} {\path{doi:10.1007/s10409-018-00837-8}}.

\bibitem{Ordoubadi2017}
M.~Ordoubadi, M.~Yaghoubi, F.~Yeganehdoust, Surface tension simulation of free surface flows using smoothed particle hydrodynamics, Sci. Iran 24 (2017) 2019--2033.
\newblock \href {https://doi.org/10.24200/sci.2017.4291} {\path{doi:10.24200/sci.2017.4291}}.

\bibitem{Hirschler2017}
M.~Hirschler, G.~Oger, U.~Nieken, D.~L. Touz{\'e}, Modeling of droplet collisions using {S}moothed {P}article {H}ydrodynamics, Int. J. Multiphase Flow 95 (2017) 175--187.
\newblock \href {https://doi.org/10.1016/j.ijmultiphaseflow.2017.06.002} {\path{doi:10.1016/j.ijmultiphaseflow.2017.06.002}}.

\bibitem{GEARA2022}
S.~Geara, S.~Martin, S.~Adami, W.~Petry, J.~Allenou, B.~Stepnik, O.~Bonnefoy, A new {SPH} density formulation for {3}d free-surface flows, Computers \& Fluids 232 (2022) 105193.
\newblock \href {https://doi.org/10.1016/j.compfluid.2021.105193} {\path{doi:10.1016/j.compfluid.2021.105193}}.

\bibitem{Fuerstenau2020}
J.-P. F{\"u}rstenau, C.~Wei{\ss}enfels, P.~Wriggers, Free surface tension in incompressible smoothed particle hydrodynamcis ({ISPH}), Comput. Mech. 65 (2020) 487--502.
\newblock \href {https://doi.org/10.1007/s00466-019-01780-6} {\path{doi:10.1007/s00466-019-01780-6}}.

\bibitem{Blank2022}
M.~Blank, P.~Nair, T.~P{\"o}schel, Modeling surface tension in {S}moothed {P}article {H}ydrodynamics using {Y}oung-{L}aplace pressure boundary condition, Comp. Meth. Appl. Mech. Engin. 406 (2023) 115907.
\newblock \href {https://doi.org/10.1016/j.cma.2023.115907} {\path{doi:10.1016/j.cma.2023.115907}}.

\bibitem{Breinlinger2013}
T.~Breinlinger, P.~Polfer, A.~Hashibon, T.~Kraft, Surface tension and wetting effects with smoothed particle hydrodynamics, J. Comput. Phys. 243 (2013) 14--27.
\newblock \href {https://doi.org/10.1016/j.jcp.2013.02.038} {\path{doi:10.1016/j.jcp.2013.02.038}}.

\bibitem{Monaghan2005}
J.~J. Monaghan, Smoothed particle hydrodynamics, Rep. Prog. Phys. 68 (2005) 1703.
\newblock \href {https://doi.org/10.1088/0034-4885/68/8/R01} {\path{doi:10.1088/0034-4885/68/8/R01}}.

\bibitem{Wendland1995}
H.~Wendland, Piecewise polynomial, positive definite and compactly supported radial functions of minimal degree, Adv. Comput. Math. (1995) 389--396\href {https://doi.org/10.1007/BF02123482} {\path{doi:10.1007/BF02123482}}.

\bibitem{Morris1997}
J.~P. Morris, P.~J. Fox, Y.~Zhu, Modeling low {Reynolds} number incompressible flows using {SPH}, J. Comput. Phys. 136 (1997) 214--226.
\newblock \href {https://doi.org/10.1006/jcph.1997.5776} {\path{doi:10.1006/jcph.1997.5776}}.

\bibitem{Cummins1999}
S.~J. Cummins, M.~Rudman, An {SPH} projection method, J. Comput. Phys. 152 (1999) 584--607.
\newblock \href {https://doi.org/10.1006/jcph.1999.6246} {\path{doi:10.1006/jcph.1999.6246}}.

\bibitem{Sleijpen1994}
G.~L.~G. Sleijpen, H.~A. Van~der Vorst, D.~R. Fokkema, {BiCGstab(l)} and other hybrid {Bi-CG} methods, Numer. Algorithms 7 (1994) 75--109.
\newblock \href {https://doi.org/10.1007/BF02141261} {\path{doi:10.1007/BF02141261}}.

\bibitem{Nair2014}
P.~Nair, G.~Tomar, An improved free surface modeling for incompressible {SPH}, Comput. Fluids 102 (2014) 304--314.
\newblock \href {https://doi.org/10.1016/j.compfluid.2014.07.006} {\path{doi:10.1016/j.compfluid.2014.07.006}}.

\bibitem{Bonet1999}
J.~Bonet, T.~S.~L. Lok, Variational and momentum preservation aspects of {Smooth} {Particle} {Hydrodynamic} formulations, Comput. Methods Appl. Mech. Eng. 180 (1999) 97--115.
\newblock \href {https://doi.org/10.1016/S0045-7825(99)00051-1} {\path{doi:10.1016/S0045-7825(99)00051-1}}.

\bibitem{Oger2007}
G.~Oger, M.~Doring, B.~Alessandrini, P.~Ferrant, An improved {SPH} method: {T}owards higher order convergence, J. Comput. Phys. 225 (2007) 1472--1492.
\newblock \href {https://doi.org/10.1016/j.jcp.2007.01.039} {\path{doi:10.1016/j.jcp.2007.01.039}}.

\bibitem{Shepard1968}
D.~Shepard, A two-dimensional interpolation function for irregularly-spaced data, in: Proceedings of the 1968 23rd ACM national conference, 1968, pp. 517--524.
\newblock \href {https://doi.org/10.1145/800186.810616} {\path{doi:10.1145/800186.810616}}.

\bibitem{Lee2008}
E.-S. Lee, C.~Moulinec, R.~Xu, D.~Violeau, D.~Laurence, P.~Stansby, Comparisons of weakly compressible and truly incompressible algorithms for the {SPH} mesh free particle method, J. Comput. Phys. 227 (2008) 8417--8436.
\newblock \href {https://doi.org/10.1016/j.jcp.2008.06.005} {\path{doi:10.1016/j.jcp.2008.06.005}}.

\bibitem{asai2012stabilized}
M.~Asai, A.~M. Aly, Y.~Sonoda, Y.~Sakai, A stabilized incompressible {SPH} method by relaxing the density invariance condition, J. Appl. Math. 2012 (2012) 139583.
\newblock \href {https://doi.org/10.1155/2012/139583} {\path{doi:10.1155/2012/139583}}.

\bibitem{Monaghan1992}
J.~J. Monaghan, Smoothed particle hydrodynamics, Annual Rev. Astron. Astrophys. 30 (1992) 543--574.
\newblock \href {https://doi.org/10.1146/annurev.aa.30.090192.002551} {\path{doi:10.1146/annurev.aa.30.090192.002551}}.

\bibitem{Gennes1985}
P.~G. de~Gennes, Wetting: statics and dynamics, Rev. Mod. Phys. 57 (1985) 827--863.
\newblock \href {https://doi.org/10.1103/RevModPhys.57.827} {\path{doi:10.1103/RevModPhys.57.827}}.

\bibitem{Violeau2015}
D.~Violeau, Fluid Mechanics and the {SPH} Method, Oxford Univ. Press, 2015.
\newblock \href {https://doi.org/10.1093/acprof:oso/9780199655526.001.0001} {\path{doi:10.1093/acprof:oso/9780199655526.001.0001}}.

\bibitem{monaghan1994simulating}
J.~J. Monaghan, Simulating free surface flows with {SPH}, J. Comput. Phys. 110 (1994) 399--406.
\newblock \href {https://doi.org/10.1006/jcph.1994.1034} {\path{doi:10.1006/jcph.1994.1034}}.

\bibitem{crespo2007}
A.~J.~C. Crespo, M.~G{\'o}mez-Gesteira, R.~A. Dalrymple, Boundary conditions generated by dynamic particles in {SPH} methods, Comput. Mater. Contin. 5 (2007) 173--184.
\newblock \href {https://doi.org/10.3970/cmc.2007.005.173} {\path{doi:10.3970/cmc.2007.005.173}}.

\bibitem{nasar2019}
A.~M.~A. Nasar, B.~D. Rogers, A.~Revell, P.~K. Stansby, Flexible slender body fluid interaction: {V}ector-based discrete element method with {E}ulerian smoothed particle hydrodynamics, Comput. Fluids 179 (2019) 563--578.
\newblock \href {https://doi.org/10.1016/j.compfluid.2018.11.024} {\path{doi:10.1016/j.compfluid.2018.11.024}}.

\bibitem{leroy2014unified}
A.~Leroy, D.~Violeau, M.~Ferrand, C.~Kassiotis, Unified semi-analytical wall boundary conditions applied to 2-d incompressible {SPH}, J. Comput. Phys. 261 (2014) 106--129.
\newblock \href {https://doi.org/10.1016/j.jcp.2013.12.035} {\path{doi:10.1016/j.jcp.2013.12.035}}.

\bibitem{adami2012}
S.~Adami, X.~Y. Hu, N.~A. Adams, A generalized wall boundary condition for smoothed particle hydrodynamics, J. Comput. Phys. 231 (2012) 7057--7075.
\newblock \href {https://doi.org/10.1016/j.jcp.2012.05.005} {\path{doi:10.1016/j.jcp.2012.05.005}}.

\bibitem{Mugele2019}
F.~Mugele, J.~Heikenfeld, Electrowetting - Fundamental Principles and Practical Applications, John Wiley \& Sons, New York, 2019.
\newblock \href {https://doi.org/10.1002/9783527412396} {\path{doi:10.1002/9783527412396}}.

\bibitem{Aalilija2020}
A.~Aalilija, C.-A. Gandin, E.~Hachem, On the analytical and numerical simulation of an oscillating drop in zero-gravity, Comput. Fluids 197 (2020) 104362.
\newblock \href {https://doi.org/10.1016/j.compfluid.2019.104362} {\path{doi:10.1016/j.compfluid.2019.104362}}.

\bibitem{Lafuma:2003}
A.~Lafuma, D.~Qu{\'e}r{\'e}, Superhydrophobic states, Nature Materials 2 (2003) 457--460.
\newblock \href {https://doi.org/10.1038/nmat924} {\path{doi:10.1038/nmat924}}.

\bibitem{Dupont2010}
J.-B. Dupont, D.~Legendre, Numerical simulation of static and sliding drop with contact angle hysteresis, J. Comput. Phys. 229 (2010) 2453--2478.
\newblock \href {https://doi.org/10.1016/j.jcp.2009.07.034} {\path{doi:10.1016/j.jcp.2009.07.034}}.

\bibitem{GOHL2018}
J.~Göhl, A.~Mark, S.~Sasic, F.~Edelvik, An immersed boundary based dynamic contact angle framework for handling complex surfaces of mixed wettabilities, International Journal of Multiphase Flow 109 (2018) 164--177.

\end{thebibliography}

\end{document}